%% file: main.tex
\documentclass{lmcs}
\pdfoutput=1
\usepackage[utf8]{inputenc}

\usepackage{lastpage}
\lmcsdoi{21}{1}{9}
\lmcsheading{}{\pageref{LastPage}}{}{}%
{Nov.~01,~2023}{Jan.~28,~2025}{}


\usepackage{xspace}
\usepackage{tabularx,environ}
\usepackage{graphicx}
\usepackage{hyperref}
\usepackage{pdfpages}

\input{myMacros.tex}
\theoremstyle{plain} %\crefname{satz}{Satz}{S\"atze}
\NewEnviron{problem}[1]{%
\begin{center}\fbox{\parbox{.9\textwidth}{%
    {\centering #1\par}%
    \parskip=1ex
    \everypar{\hangindent=1em}%
    \BODY
}}
\end{center}}
\newcommand{\revc}[1]{\textcolor{black}{#1}}  

\begin{document}

\title{Diversity of Answers to Conjunctive Queries}
\titlecomment{
{\lsuper*}This is an extended and enhanced version of an article published at ICDT 2023 \cite{unsereKonferenzversion}}
\thanks{{\lsuper*}This work was supported by the Austrian Science Fund (FWF) project P30930-N35 and by the Vienna Science and Technology Fund (WWTF) [10.47379/ICT2201].}	%optional

% affiliations are numbered automatically with a, b, c (see below)
% use the optional argument to indicate the affiliation(s) of each author
% omit the argument if there is only one author, or only one affiliation
\author[T.~Merkl]{Timo Camillo Merkl\lmcsorcid{0000-0002-1825-0097}}[a]
\author[R.~Pichler]{Reinhard Pichler\lmcsorcid{0009-0003-7206-2518}}[a,b]
\author[S.~Skritek]{Sebastian Skritek\lmcsorcid{0000-0003-3054-7683}}[a]

% affiliation 1 (automatically numbered a)
\address{TU Wien, Vienna, Austria}%optional
% write emails for all authors having that affiliation
\email{timo.merkl@tuwien.ac.at, reinhard.pichler@tuwien.ac.at, sebastian.skritek@tuwien.ac.at}  %optional

% affiliation 2 (automatically numbered b)
% \address{University 2, address2}	%optional
% \email{name2@email2}  %optional

%% etc.

%% required for running head on odd and even pages, use suitable
%% abbreviations in case of long titles and many authors:

%%%%%%%%%%%%%%%%%%%%%%%%%%%%%%%%%%%%%%%%%%%%%%%%%%%%%%%%%%%%%%%%%%%%%%%%%%%

%% the abstract has to PRECEDE the command \maketitle:
%% be sure not to issue the \maketitle command twice!

\begin{abstract}
  \noindent Enumeration problems aim at outputting, without repetition,
    the set of solutions to a given problem instance. 
    However, outputting the entire 
    solution set may be prohibitively expensive if it is too big. 
    In this case, outputting a small,  sufficiently diverse subset of the solutions 
    would be preferable.
    This leads to the Diverse-version of 
    the original enumeration problem, where the goal is to achieve 
    a certain level d of diversity by selecting k solutions. 
    In this paper, we look at the Diverse-version of the query answering problem for Conjunctive Queries and extensions thereof. That is, we study the problem if it is possible to 
    achieve a certain level d of diversity by selecting k answers to the given query and, 
    in the positive case, to actually compute such k answers.
\end{abstract}

\maketitle

%% start the paper here:
\section{Introduction}
\label{sec:introduction}

The notion of {\it solutions} is ubiquitous in Computer Science
and there are many ways of defining computational problems to
deal with them. Decision problems, for instance, may ask if the 
set of solutions is non-empty or test for a given candidate if it 
indeed is a solution. Search problems aim at finding a concrete solution 
and counting problems aim at determining  the number
of solutions. In recent time, {\em enumeration problems}, which 
aim at outputting, without repetition,
the set of solutions to a given problem instance 
have gained a lot of interest, 
which is, for instance, witnessed by two recent Dagstuhl seminars on this topic
\cite{DBLP:journals/dagstuhl-reports/BorosKPS19,DBLP:journals/dagstuhl-reports/FernauGS18}.
Also in the Database Theory community, enumeration problems have 
played a prominent role on the research agenda recently, see e.g., 
\cite{DBLP:conf/pods/AmarilliJMR22,DBLP:conf/pods/KobayashiKW22,DBLP:conf/pods/LutzP22,%
DBLP:conf/icdt/Deng0023,DBLP:conf/icdt/MunozR23,DBLP:conf/pods/CarmeliS23}.
Here, the natural problem to consider is 
query answering with the answers to a given query constituting the 
``solutions'' to this problem.

It is well known that even seemingly simple problems, 
such as answering an acyclic Conjunctive Query, can have a huge number of solutions. Consequently, specific notions of tractability 
were introduced right from the beginning of research on 
enumeration problems~\cite{DBLP:journals/ipl/JohnsonP88} 
to separate the computational
intricacy of a problem from the mere size of the solution space. 
However, even with these refined notions of tractability, 
the usefulness of flooding the user with tons of solutions (many of them possibly differing only minimally) may be questionable. 
If the solution space gets too big, it would be more useful to provide an overview by outputting a ``meaningful'' subset of the solutions. One way of 
pursuing this goal is to randomly select solutions (also known as ``sampling'')
as was, for instance, done
in~\cite{DBLP:conf/pods/ArenasCJR19,DBLP:conf/pods/CarmeliZBKS20}. 
In fact, research on sampling has a long tradition in the Database community 
\cite{DBLP:journals/debu/ChaudhuriM99} -- 
above all with the goal of supporting more accurate cardinality estimations \cite{DBLP:conf/cidr/LeisRGK017,DBLP:journals/tods/LiWYZ19,DBLP:conf/sigmod/ZhaoC0HY18}.   

A different approach to providing a ``meaningful'' subset of the solution space 
aims at outputting a small {\em diverse} subset of the solutions. 
This approach has enjoyed considerable popularity in the Artificial Intelligence community 
\cite{DBLP:journals/ai/BasteFJMOPR22,DBLP:journals/tplp/EiterEEF13,%
DBLP:conf/sat/Nadel11} -- 
especially when dealing with Constraint Satisfaction Problems (CSPs) 
\cite{DBLP:conf/aaai/HebrardHOW05,DBLP:conf/aaai/IngmarBST20,%
DBLP:conf/ijcai/PetitT15}.
For instance, consider a variation of the car dealership example from \cite{DBLP:conf/aaai/HebrardHOW05}.
Suppose that $I$ models the preferences of a customer and $\mathcal{S}(I)$ are all cars that match these restrictions.
Now, in a large dealership, presenting all cars in $\mathcal{S}(I)$ to the customer would be infeasible. 
Instead, it would be better to go through a rather small list of cars that are significantly different from each other. 
With this, the customer can point at those cars which the further discussion with the clerk should concentrate on.

Due to the inherent hardness of achieving the maximal possible diversity \cite{DBLP:conf/aaai/HebrardHOW05}, the Database community -- apart from limited exceptions \cite{DBLP:journals/tods/DengF14} -- focused on heuristic and approximation methods to find diverse solutions (see \cite{DBLP:journals/kais/ZhengWQLG17} for an extensive survey).
Also, there, diversification is usually treated as a post-processing task that is applied to a set of solutions after materializing it.

The goal of our work is therefore to broaden the understanding of the theoretical boundaries of diverse query answering and develop complementary exact algorithms. 
More specifically, we want to analyze diversity problems 
related to answering Conjunctive Queries (CQs) and extensions thereof. 
As pointed out in~\cite{DBLP:conf/aaai/IngmarBST20},
to formalize the problems we are thus studying, we, first of all, have to 
fix a notion of {\em distance} between any two solutions and an 
{\em aggregator} to combine pairwise distances to a
{\em diversity measure} for a set of solutions. For the distance between 
two answer tuples, we will use the Hamming distance throughout this paper, 
that is, counting the number of positions on which two tuples differ. 
Our developed techniques naturally extend to weighted Hamming distances, i.e., where each attribute is assigned a constant weight and we sum over the weights of those attributes the two answers differ on.
However, for the sake of simplicity, we stick to the unweighted version in our discussion.
As far as the choice of an aggregator $f$ is concerned, we impose the general restriction 
that it must be computable in polynomial time\footnote{The reason for this restriction is simply to not dilute our discussion of the complexity of the diversity problem with the inherent complexity of computing the diversity of a set of answers itself.
If we are \revc{content} with higher complexity classes, like $\fpt$, the time required to compute the aggregator may also take longer, e.g., $\fpt$ time.}.
As will be detailed below, we will sometimes also consider more restricted cases of aggregators.
Formally, for a class $\calQ$ of queries and diversity measure $\delta$ that maps 
$k$ answer tuples to an aggregated distance,  
we will study the 
following problem \DiverseQ ($\delta$ is fixed and not part of the input): 

\begin{problem}{\DiverseQ}
    Input: A database instance $I$, query $Q \in \calQ$, and integers $k$ and $d$.
    
    Question: Do there exist pairwise distinct answers 
    $\gamma_1,\dots,\gamma_k$ to $Q$ over $I$ such that $\delta(\gamma_1,\dots,\gamma_k)\geq d$?
\end{problem}

\noindent
That is, we ask if a certain level $d$ of diversity 
can be achieved by choosing $k$ pairwise distinct answers to 
a given query $Q$ over the database instance 
$I$. We refer to $\{\gamma_1,\dots,\gamma_k\}$ as the desired {\em diversity set}.
\revc{In the literature, one can find examples of duplicates being allowed and disallowed in the diversity set.
In the present work, we disallow duplicates if not mentioned otherwise since it seems counter-intuitive to expect a user to get a broader picture of the solution space when 
being presented with the same element multiple times.
However, we note that all results in this paper remain the same no matter whether we exclude duplicates 
or not 
-- with one single exception: for the query complexity of the \DiverseACQsum (formally defined below) problem, 
we manage to show $\ptime$-membership if duplicates are allowed (see Theorem~\ref{theorem:querycomplexitypolytime}), but only $\fpt$-membership if duplicates are excluded (see Theorem~\ref{theorem:DiverseACQsum}).}

As far as the notation is concerned, we will denote the Hamming distance between 
two answers $\gamma$, $\gamma'$ by $\Delta(\gamma, \gamma')$.
With diversity measure $\delta$, we denote the aggregated Hamming distances of all pairs of $k$ answer
tuples for an arbitrary, polynomial-time computable aggregate function $f$. 
That is, 
let 
$f\colon \bigcup_{k\geq 1} \mathbb{N}^{\frac{k(k-1)}{2}} \rightarrow \mathbb{R}$
and let $d_{i,j} = \Delta(\gamma_i, \gamma_j)$ for $1\leq i < j \leq k$. 
Then we 
define $\delta(\gamma_1, \dots, \gamma_k): = f((d_{i,j})_{1\leq i < j \leq k})$.
Sometimes it will be necessary to restrict our attention to concrete aggregators or concrete classes of aggregators.
To that end, we write $\deltasum$ if the aggregator $f$ is the sum, 
$\deltamon$ if the aggregator $f$ 
is a monotone function, i.e., 
$f(d_1, \dots, d_N)
\leq 
f(d'_1, \dots, d'_N)$
whenever 
$d_i \leq d'_i$
holds for every $i \in \{1, \dots, N\}$ with 
$N = \frac{k(k-1)}{2}$,
and $\deltastrmon$ if the aggregator $f$ is weakly strictly monotone (ws-monotone) in the sense that
$f(d_1, \dots, d_N)
< 
f(d, \dots, d)$
whenever 
$d_i \leq d$
holds for every $i \in \{1, \dots, N\}$ and at least one $d_i < d$.
Note that most natural measures of diversity are ws-monotone, e.g., aggregating via sum or min but may not be strictly monotone, e.g., min.
To emphasize the class of diversity measures in question, we denote the corresponding diversity problems by 
\DiverseQsum, \DiverseQmon, and \DiverseQstrmon, respectively.

When we prove upper bounds on the complexity of several variations of the \DiverseQ problem
(in the form of membership in some favorable complexity class), we aim at the most general 
setting, i.e., membership for all polynomial-time computable aggregation functions. 
However, in some cases, the restriction to \DiverseQsum or \DiverseQmon
will be needed 
in order to 
achieve the desired upper bound on the complexity. In contrast, to prove lower bounds (in the form of hardness results), 
we consider arbitrary ws-monotone diversity measures, i.e., \DiverseQstrmon.
This means that our hardness results do not just hold for a particular diversity measure but for \textit{all} fixed, ws-monotone diversity measures.

We will analyze the \DiverseQ problem for several query 
classes $\calQ$ -- starting with the class $\CQ$ of Conjunctive Queries and 
then extending our studies to the classes $\UCQ$ and $\CQneg$ of unions of CQs and 
CQs with negation. In one case, we will also look at the class $\FO$ of all first-order queries.
Recall that, for combined complexity and query complexity, even the question, if an answer tuple exists at all, is $\np$-complete for CQs~\cite{DBLP:conf/stoc/ChandraM77}.
We therefore mostly restrict our study to acyclic CQs (ACQs, for short) 
with the corresponding query classes
$\ACQ$ and $\UACQ$, allowing only ACQs and unions of ACQs, respectively.
For CQs with negation, query answering remains 
$\np$-complete even if we only allow ACQs 
\revc{\cite{DBLP:journals/tcs/Lanzinger23}}. 
Hence, for $\CQneg$ 
we have to impose a different restriction. We thus restrict ourselves to CQs with 
bounded treewidth. Finally note that, even if we have formulated \DiverseQ as a decision problem, 
we also care about actually computing $k$ solutions in case of a yes-answer.

We aim at a thorough complexity analysis of the \DiverseQ problem from 
various angles. For the most part, we consider the problem parameterized by the 
size $k$ of the diversity set.
In the non-parameterized case 
(i.e., if $k$ is simply part of the input) we assume $k$ 
to be given in unary representation. This assumption is motivated by the fact that for binary representation of $k$, the size $k$ of the diversity set
can be exponentially larger than the input: this contradicts the spirit of the diversity approach which aims at outputting a {\em small} (not an exponentially big) number of diverse solutions. As is customary in the 
Database world, we will distinguish combined, query, and data complexity.

\medskip

\noindent
\subsubsection*{\bf Summary of results.} 
\nopagebreak

\begin{itemize}
\item 
We start our analysis of the \DiverseQ problem with the class of ACQs
and study data complexity, query complexity, and combined complexity. With the size $k$ of the diversity set
as the parameter, we establish $\xp$-membership for combined complexity, which is strengthened to 
$\fpt$-membership for data complexity. The $\xp$-membership of combined complexity is 
complemented by a $\wone$-lower bound of the \DiverseACQstrmon problem. 
For the non-parameterized case, we show that even the data complexity is $\np$-hard.
\item 
The $\fpt$-result of data complexity is easily extended to unions of ACQs. 
Actually, it even holds for arbitrary FO-queries. 
However, rather
surprisingly, we show that the combined complexity and even query complexity of the 
\DiverseUACQstrmon problem is $\np$-complete \revc{even when only looking for a pair of diverse answers. 
That is,} the hardness still holds 
if the size $k$ of the diversity set
is 2 and the UACQs are restricted to unions of 2 ACQs. 
\item 
Finally, we study the  \DiverseQ problem for the class $\CQneg$. 
As was mentioned above, the restriction to ACQs is not even enough to make the query answering
problem tractable. We, therefore, study the \DiverseCQn problem by allowing only classes of 
CQs of bounded treewidth. The picture is then quite similar to the 
\DiverseACQ problem, featuring  analogous $\xp$-membership, $\fpt$-membership, 
$\wone$-hardness, and  $\np$-hardness results.
\end{itemize}

\noindent
{\bf Structure.} We present  some basic definitions and results in 
Section~\ref{sec:preliminaries}. In particular, we will formally 
introduce all concepts of parameterized complexity (complexity classes, 
reductions) relevant to our study.
We then analyze various variants of the \DiverseQ problem, where
$\calQ$ is the class of CQs in Section~\ref{sec:CQs},
the class of unions of CQs in Section~\ref{sec:UCQs}, and
the class of CQs with negation in Section~\ref{sec:CQneg},
respectively.
Some conclusions and directions for future work are given in 
Section~\ref{sec:conclusion}.

\section{Preliminaries}
\label{sec:preliminaries}

{\em Basics.}
We assume familiarity with relational databases. For basic notions such as 
schema, (arity of) relation symbols, relations, (active) domain,  etc., 
the reader is referred to any database textbook, 
e.g., \cite{DBLP:books/aw/AbiteboulHV95}. 
A CQ is a first-order formula of the form
\[ Q(X):= \exists Y \bigwedge_{i=1}^\ell A_i,\] with 
free variables $X= (x_1,\dots,x_m)$ and 
bound variables $Y= (y_1, \dots, y_n)$ such that 
each $A_i$ is an atom with variables from 
$x_1,\dots,x_m, y_1, \dots, y_n$. 
An answer to such a CQ $Q(X)$ over a database instance 
(or simply ``database'', for short) $I$ 
is a mapping $\gamma \colon X \rightarrow \dom(I)$
which can be extended to a mapping 
$\bar \gamma \colon (X \cup Y) \rightarrow \dom(I)$
such that instantiating each variable $z \in (X \cup Y)$
to $\bar \gamma (z)$ sends each 
atom $A_i$ into the database $I$. 
We write $\dom(I)$ to denote the (finite, active) domain of $I$. 
By slight abuse of notation, we also refer to the 
tuple $\gamma(X)=(\gamma (x_1),\dots, \gamma (x_m))$ as an answer (or an answer tuple).
A UCQ is a disjunction $\bigvee_{i=1}^{N} Q_i(X)$, where all $Q_i$'s are CQs with the same  
free variables. The set of answers of a UCQ is the union of the answers of its CQs.
In a CQ with negation, we allow the $A_i$'s to be either (positive) atoms or
literals (i.e., negated atoms) satisfying a safety condition, i.e., every variable has to occur in 
some positive atom. An answer 
to a CQ with negation $Q(X)$ over a database $I$ has to satisfy the 
condition that each positive atom is sent to an atom in the database while each negated atom is not.
The set of answers to a query $Q$ over a database $I$ is 
denoted by $Q(I)$.

For two mappings $\alpha$ and $\alpha'$ defined on variable sets 
$Z$ and $Z'$, respectively, we write $\alpha \cong \alpha'$ to denote that 
the two mappings coincide on all variables in $Z \cap Z'$. If this is the case, 
we write $\alpha \cap \alpha'$ and $\alpha \cup \alpha'$ to denote the mapping obtained by
restricting $\alpha$ and $\alpha'$ to their common domain or by combining them
to the union of their domains, respectively. 
That is, $(\alpha \cap \alpha')(z) = \alpha (z)$ for every $z \in Z \cap Z'$ and
$(\alpha \cup \alpha')(z)$ is either  $\alpha (z)$ if $z \in Z$ or $\alpha'(z)$ otherwise.
For another variable set $X$ and $z \in Z$, we write $\alpha|_{X}$ and $\alpha|_{z}$ for the \revc{mappings} 
resulting from the restriction of $\alpha$ to the set $X\cap Z$ or the singleton $\{z\}$, respectively. Also, the Hamming distance between two mappings can be restricted to a 
subset of the positions (or, equivalently, of the variables): 
by  $\Delta_X (\alpha,\alpha')$ we denote the number of variables in $X$
on which $\alpha$ and $\alpha'$ differ.

\medskip

\noindent
{\em Acyclicity and widths.}
In a landmark paper~\cite{DBLP:conf/vldb/Yannakakis81}, Yannakakis
showed that query evaluation is tractable (combined complexity)
if restricted to {\em acyclic}
CQs. A CQ is acyclic if it has a {\em join tree}.  
Given a CQ $Q(X):= \exists Y \bigwedge_{i=1}^\ell A_i$
with $At(Q(X)) = \{A_i : 1 \leq i \leq \ell\}$,
a join tree of $Q(X)$ is a triple $\langle T, \lambda, r \rangle$
such that $T=(V(T),E(T))$ is a rooted tree with root $r$ and 
$\lambda \colon V(T) \rightarrow At(Q(X))$ is a node labeling function
that satisfies the following properties:
\begin{enumerate}
    \item The labeling $\lambda$ is a bijection.
    \item For every $v\in X \cup Y$, the set 
    $T_v = \{t\in V(T) : v$ occurs in $\lambda(t)\}$ induces a subtree~$T[T_v]$ of $T$.
\end{enumerate}
As is common, for a graph $T = (V(T), E(T))$ and $H \subseteq V(T)$, we use $T[H]$ to denote
the subgraph of $T$ induced by $H$, i.e.\ the subgraph consisting of $H$ and all edges in 
$E(T)$ between nodes in $H$.

Testing if a given CQ is acyclic and, in case of a yes-answer, constructing 
a join tree is feasible in polynomial time by the GYO-algorithm, named after the authors of  \cite{report/toronto/Gra79,DBLP:conf/compsac/YuO79}.

Another approach to making CQ answering tractable is 
by restricting the {\em treewidth} ($\tw$), 
which is defined via {\em tree decompositions}~\cite{DBLP:journals/jct/RobertsonS84}. Treewidth does not generalize acyclicity, 
i.e., a class of acyclic CQs can have unbounded $\tw$.
We consider $\tw$ here only for CQs with negation.
Let  $Q(X):= \exists Y \bigwedge_{i=1}^\ell L_i$,
be a CQ with negation, i.e., each $L_i$ is a 
(positive or negative) literal. Moreover, 
let  $\var(L_i)$ denote the variables
occurring in $L_i$. 
A tree decomposition of $Q(X)$ is a 
triple $\langle T, \chi, r \rangle$
such that $T=(V(T),E(T))$ is a rooted tree with root $r$ and 
$\chi \colon V(T) \rightarrow 2^{X\cup Y}$ 
is a node labeling function 
with  the following properties:
\begin{enumerate}
    \item For every $L_i$, there exists a node $t \in V(T)$ with $\var(L_i) \subseteq \chi(t)$.
    \item For every $v\in X \cup Y$, the set 
    $T_v = \{t\in V(T) : v \in \chi(t)\}$ induces a subtree~$T[T_v]$ of $T$.
\end{enumerate}
The property (2) is called the connectedness condition for join trees and tree decomposition. 
The sets $\chi(t)$  of variables are referred to as ``bags'' of the 
tree decomposition $T$. 
The width 
of a tree decomposition is defined as $\max_{t\in V(T)} (|\chi(t)| -1)$.
The treewidth of a CQ with negation $Q$ is the minimum width of all tree decompositions of $Q$.
For fixed $\omega$, it is feasible in linear time w.r.t.\ the size of the query $Q$ to decide if
$\tw(Q) \leq \omega$ holds and, in case of a yes-answer, to actually compute a tree decomposition  
of width $\leq \omega$~\cite{DBLP:journals/siamcomp/Bodlaender96}.

\revc{
Tree decompositions can be extended to hypertree decompositions (HDs)
\cite{DBLP:journals/jcss/GottlobLS02},
generalized hypertree decompositions (GHDs) \cite{DBLP:journals/ejc/AdlerGG07},
and fractional hypertree decompositions (FHDs) \cite{DBLP:journals/talg/GroheM14}
by defining an integral edge cover number (in case of HDs and GHDs) 
or a fractional edge cover number
(in case of FHDs) for each bag. Then the width of such a decomposition is the 
maximum size of all edge cover numbers in the HD, GHD, or FHD. 
Analogously to treewidth, the hypertree width $\hw(Q)$, 
generalized hypertree width $\ghw(Q)$, and fractional hypertree width
$\fhw(Q)$ of a given query $Q$ is defined as the minimum width attainable 
over all HDs, GHDs, or FHDs of $Q$, respectively. Acyclic queries are the ones
with $\hw = \ghw = \fhw =1$.
}

\revc{
Note that bounded treewidth and acyclicity are incomparable. Indeed, in case of 
unbounded arity, even the class of single-atom queries (clearly, all such queries are trivially acyclic) has unbounded treewidth. On the other hand, for instance, the 
triangle query and its generalization to any cycles has constant treewidth 2, but these queries are, of course, not acyclic. However, the classes of queries with bounded $\hw$, $\ghw$, or $\fhw$ properly contain the classes of queries with 
bounded $\tw$. Moreover, query evaluation of queries with bounded 
$\hw$, $\ghw$, or $\fhw$ can be 
reduced in polynomial time to query evaluation of acyclic queries. Essentially, 
the relations of the resulting acyclic query are obtained by carrying out the 
joins corresponding to each edge cover.}

\medskip
\noindent
{\em Complexity.} 
We follow the categorization of the complexity of database tasks introduced in~\cite{DBLP:conf/stoc/Vardi82} and distinguish
combined/query/data complexity of the 
\DiverseQ problem. That is, for data complexity, we consider the query
$Q$ as arbitrarily chosen but fixed, while for query complexity, the 
database instance $I$ is considered fixed. In case of combined complexity, both the query and the database are considered as variable parts of the input.

We assume familiarity with the fundamental complexity classes $\ptime$ (polynomial time) and $\np$ (non-deterministic polynomial time).
We  study the \DiverseQ problem primarily from a 
parameterized complexity perspective~\cite{DBLP:series/mcs/DowneyF99}. 
An instance of a {\em parameterized problem\/} is given as a pair $(x,k)$, where $x$ is the actual problem instance and $k$ is a parameter -- usually a non-negative integer. The effort for 
solving a parameterized problem is measured by a function that depends on both, the size $|x|$ of the instance and the value $k$ of the parameter. The asymptotic
worst-case time complexity is thus specified as $\O(f(n,k))$ 
with $n = |x|$. 

The parameterized analogue of {\em tractability\/} 
captured by the class $\ptime$ is {\em fixed-parameter tractability} 
captured by the class $\fpt$ of fixed-parameter tractable problems.
A problem is in $\fpt$, 
if it can be solved in time $\O(f(k) \cdot n^c)$ for some computable function $f$ and constant~$c$. In other words, the running time only depends polynomially on the size of the instance, while a possibly exponential explosion is confined to the parameter. In particular, if for a class of instances, the parameter $k$ is bounded by a constant, then $\fpt$-membership means that 
the problem can be solved in polynomial time. This also applies to 
problems in the slightly less favorable complexity class $\xp$, 
which contains the problems solvable in time $\O(n^{f(k)})$.

Parameterized complexity theory also comes with its own version of 
reductions (namely ``$\fpt$-reductions'') and hardness theory based on 
classes of fixed-parameter intractable problems. 
An $\fpt$-reduction from a parameterized problem $P$ to another
parameterized problem $P'$  maps every instance $(x,k)$ of $P$ to an 
equivalent instance $(x',k')$ of $P'$, such that~$k'$  only depends on $k$ (i.e., independent of $x$) and the computation of $x'$ is in $\fpt$ (i.e., 
in time $\O(f(k) \cdot |x|^c)$ for some computable function $f$ and constant $c$).
For {\em fixed-parameter intractability\/}, the 
most prominent class 
is $\wone$. It has several equivalent definitions, for instance, $\wone$ is  the class
of problems that allow for an $\fpt$-reduction to the \IndependentSet problem parameterized by the desired size $k$ of an independent set. 
We have \mbox{$\fpt \subseteq \wone \subseteq \xp$}. 
It is a generally accepted assumption in parameterized complexity theory that 
$\fpt \neq \wone$ holds -- similar but slightly stronger than
the famous $\ptime \neq \np$ assumption in classical complexity theory, i.e., 
$\fpt \neq \wone$ implies $\ptime \neq \np$, but not vice versa.

\section{Diversity of Conjunctive Queries}
\label{sec:CQs}

\subsection{Combined and Query Complexity}
\label{sec:CQ-combined}

We start our study of the \DiverseACQ problem by considering the 
combined complexity and then, more specifically, the query complexity. 
We will thus present our basic algorithm in Section~\ref{sec:CQ-algorithm}, which allows us to establish 
the $\xp$-membership of this problem. We will then 
prove $\wone$-hardness in 
Section~\ref{sec:wonehardness} and present 
some further improvements of the basic algorithm 
in Section~\ref{sec:speedups}.

\subsubsection{Basic Algorithm}
\label{sec:CQ-algorithm}

Our algorithm for solving \DiverseACQ is based on a dynamic programming idea analogous to 
the Yannakakis algorithm. 
Given a join tree $\langle T, \lambda, r \rangle$
and database $I$,  the Yannakakis algorithm decides in a bottom-up traversal of $T$ 
at each node $t \in V(T)$ and for each answer~$\alpha$ to the 
single-atom query $\lambda(t)$ if it can be extended to an answer to the 
CQ consisting of all atoms labeling the nodes in the complete subtree $T'$ rooted at $t$.
It then stores this (binary) information by either keeping or dismissing $\alpha$.
Our algorithm for \DiverseACQ implements a similar idea.
At its core, it stores {\em $k$-tuples} $(\alpha_1, \dots, \alpha_k)$ of answers 
to the single-atom query~$\lambda(t)$, 
each $k$-tuple describing a set of (partial) diversity sets.
We extend this information by the various vectors 
$(d_{i,j})_{1 \leq i < j \leq k}$ of Hamming distances that 
are attainable by possible extensions $(\gamma_1, \dots, \gamma_k)$
to the CQ consisting of the atoms labeling the nodes in $T'$.

In the following, we consider an \acq $Q(X) := \exists Y \bigwedge_{i=1}^{\ell} A_i$ where each atom is of the
form $A_i = R_i(Z_i)$ for some relation symbol $R_i$
and variables $Z_i \subseteq X \cup Y$.
For an atom $A = R(Z)$ and a database instance $I$, 
define $\sols{A}{I}$ as the set of mappings 
$\{\alpha \colon Z \rightarrow \dom(I) \setdefsep \alpha(Z) \in R^I\}$.
We extend the definition to sets (or conjunctions) $\psi(Z)$ of atoms 
$A_i(Z_i)$ with $Z_i \subseteq Z$.
Then $\sols{\psi}{I}$ is the set of mappings 
$\{\alpha \colon Z \rightarrow \dom(I) \setdefsep \alpha(Z_i) \in R_i^I
\text{ for all } R_i(Z_i) \in \psi(Z)\}$.
Let $\langle T, \lambda, r \rangle$ be a join tree. 
For a subtree $T'$ of $T$ we define $\lambda(T') = \{\lambda(t) \setdefsep t \in V(T')\}$
and, by slight abuse of notation, we write 
$\sols{t}{I}$ and $\sols{T'}{I}$ instead of 
$\sols{\lambda(t)}{I}$ and $\sols{\lambda(T')}{I}$.
Now consider $T'$ to be a subtree of $T$ with root $t$.
For tuples 
\[e \in \{ (\alpha_1, \dots, \alpha_k, (d_{i,j})_{1\leq i < j \leq k} )
\setdefsep \alpha_1, \dots, \alpha_k \in \sols{t}{I}, d_{i,j} \in 
\{0, \dots, |X|\} \text{ for } 1 \leq i < j \leq k\},\]
we define 
\begin{align*}
\ext{T'}(e) = \{(\gamma_1, \dots, \gamma_k) \setdefsep \gamma_1, \dots, \gamma_k \in 
\sols{T'}{I} \text{ s.t.\ } & \alpha_i \cong \gamma_i \text{ for } 1 \leq i \leq k \text{ and } \\
& \Delta_X(\gamma_i, \gamma_j) = d_{i,j} \text{ for } 1 \leq i < j \leq k\}.
\end{align*}

Intuitively, the algorithm checks for each such tuple $e$ whether there exist extensions~$\gamma_i$ of~$\alpha_i$ that 
\begin{enumerate}
    \item are solutions to the subquery induced by $T'$, and 
    \item exhibit $d_{i,j}$ as their pairwise Hamming distances. If this is the case, the
tuple $e$ is kept, otherwise, $e$ is dismissed. 
\end{enumerate}
In doing so, the goal of the algorithm 
is to compute sets $D_{T'}$ that contain exactly those $e$ with
$\ext{T'}(e) \neq \emptyset$. Having computed $D_{T}$ (i.e., for the whole join tree),
\DiverseACQ can now be decided by computing for each $e \in D_{T}$ the diversity measure
from the values $d_{i,j}$.

To do so, in a first phase, at every node $t \in V(T)$, we need to compute and store the set~$D_{T'}$
(for $T'$ being the complete subtree rooted in $t$).
We compute this set by starting with some set $D_t$ and updating it until eventually, it is
equal to $D_{T'}$. In addition, to every entry $e$ in every set $D_{t}$, we maintain a
set $\rho_{D_{t}}(e)$ containing provenance information on~$e$.
Afterwards, in the recombination phase, the sets $D_{T'}$ and $\rho_{D_{t}}(\cdot)$ are used
to compute a diversity set with the desired diversity -- if such a set exists.

\begin{algo} \label{algo:one}
Given $Q(X)$, $I$, $\langle T, \lambda, r \rangle$, $k$, $d$, and a diversity measure $\delta$ defined via some
aggregate function $f$, the first phase proceeds in three main steps:
\begin{itemize}
    \item \textbf{Initialization:} In this step, for every node $t \in V(T)$, initialize the set $D_t$ as 
    \[
        D_t = \{(\alpha_1, \dots, \alpha_k, (d_{i,j})_{1\leq i < j \leq k}) \setdefsep
            \alpha_i \in \sols{t}{I}, d_{i,j} = \Delta_X(\alpha_i, \alpha_j)\}.
    \]
            That is, $D_t$ contains one entry for every combination
            $\alpha_1, \dots, \alpha_k \in \sols{t}{I}$, and each value $d_{i,j}$ 
            ($1 \leq i < j \leq k$) is the Hamming distance of the mappings
            $\alpha_i|_{X}$ and $\alpha_j|_{X}$.

            For every $e \in D_t$, initialize $\rho_{D_t}(e)$ as the empty set.
            
        Finally, set the status of all
        non-leaf nodes in $T$ to ``not-ready'' and the status of all leaf nodes to ``ready''. 

    \item \textbf{Bottom-Up Traversal:} 
        Then repeat the following action until no ``not-ready'' node is left:
        Pick any ``not-ready'' node $t$ that has at least one ``ready'' child node $t'$.
        Update $D_t$ to \revc{$\Dnew_{t}$} as
        \begin{align*}
            \revc{\Dnew_{t}} = \{(\alpha_1, \dots, \alpha_k, 
            (\bar{d}_{i,j})_{1\leq i < j \leq k}) 
                \setdefsep 
                    {}& (\alpha_1, \dots, \alpha_k, (d_{i,j})_{1\leq i < j \leq k} ) \in D_{t}, \\
                    & (\alpha'_1, \dots, \alpha'_k, (d'_{i,j})_{1\leq i < j \leq k} ) \in D_{t'}, \\
                    & \alpha_i \cong \alpha'_i \text{ for } 1 \leq i \leq k, \\
                    & \bar{d}_{i,j} = d_{i,j} + d'_{i,j} - 
                        \Delta_{X}(\alpha_i \cap \alpha'_i, \alpha_j \cap \alpha'_j) \\
                    & \hspace{12em} \text{ for } 1 \leq i < j \leq k
                    \}.
        \end{align*}
        Expressed in a more procedural style: Take every entry $e \in D_{t}$ and compare it
        to every entry 
        $e' \in D_{t'}$. If the corresponding mappings $\alpha_i \in D_t$ and $\alpha'_i \in D_{t'}$
        agree on the shared variables, the new set \revc{$\Dnew_{t}$} contains an entry $\bar{e}$ with the mappings
        $\alpha_i$ from $e$ and the Hamming distances computed from $e$ and $e'$ as described above.

        Set $\rho_{\revc{\Dnew_t}}(\bar{e}) = \rho_{D_t}(e) \cup \{ (t', e') \}$. If the same entry $\bar{e}$
        is created from different pairs $(e, e')$, choose an arbitrary one of them for
        the definition of $\rho_{\revc{\Dnew_t}}(\bar{e})$.
        
        Finally, change the status of $t'$ from ``ready'' to ``processed''. The status of $t$
        becomes ``ready'' if the status of all its child nodes is ``processed'' and remains
        ``not-ready'' otherwise.
    
    \item \textbf{Finalization:} Once the status of root $r$ is ``ready'', remove all 
        $(\alpha_1, \dots, \alpha_k, (d_{i,j})_{1\leq i < j \leq k} )$ $\in D_r$ where 
        $f( (d_{i,j})_{1\leq i < j \leq k} ) < d$. To ensure that all answers in the diversity set
        are pairwise distinct, also remove all entries where $d_{i,j} = 0$ for some
        $(i,j)$ with $1 \leq i < j \leq k$.

        If, after the deletions, $D_r$ is empty, then there exists no diversity set of size
        $k$ with a diversity of at least $d$. Otherwise,  at least one such 
        diversity set exists.
    \end{itemize}
\end{algo}
\noindent
Clearly, the algorithm is well-defined and terminates. 
In the following theorem,  
we show that the algorithm decides \DiverseACQ and we give an upper bound on its running time.

\begin{thm}\label{theorem:acqXP}
 The \DiverseACQ problem is in $\xp$ \revc{in} combined complexity
 when parameterized by the size $k$ of the diversity set.
 More specifically, for an ACQ $Q(X)$, a database~$I$, and integers $k$ and $d$,
 \autoref{algo:one} decides the \DiverseACQ problem in time 
 $\O\big(|R^I|^{2k} \cdot (|X| + 1)^{k(k-1)} $ ${}\cdot \pol(|Q|, k)\big)$ where $R^I$ is the relation
 from $I$ with the most tuples and $\pol(|Q|,k)$ is a polynomial in  $|Q|$ and $k$.
\end{thm}

\noindent
This result is a consequence of the correctness of \autoref{algo:one}.
We show both, the correctness of the algorithm and \autoref{theorem:acqXP}
using a sequence of lemmas, discussed and proven next. 
Consider an \acq $Q(X)$ with join tree $\langle T, \lambda, r\rangle$, a database instance $I$, and
integers $k$ (the number of elements in the diversity set) and $d$ (the required diversity).
For a subtree $T'$ of $T$ with root \revc{$t$}, \revc{let $D_{T'}$} be the set of tuples 
\begin{align*}
e  \in\{(\alpha_1, \dots, \alpha_k, (d_{i,j})_{1 \leq i < j \leq k}) \setdefsep \mbox{}
    & \alpha_1, \dots, \alpha_k \in \sols{t}{I} \text{ and } \\
    & d_{i,j} \in \{0, \dots, |X|\} \text{ for } 1 \leq i < j \leq k \}
\end{align*}
such that the set
\begin{align*}
    \ext{T'}(e) = \{(\gamma_1, \dots, \gamma_k) \setdefsep 
        \gamma_1, \dots, \gamma_k \in  \sols{T'}{I} \text{ s.t.\ } &{}
                \alpha_i \cong \gamma_i \text{ for } 1 \leq i \leq k \text{ and } \\
                & \Delta_X(\gamma_i, \gamma_j) = d_{i,j} \text{ for } 1 \leq i < j \leq k\}
\end{align*}
is not empty.
To prove the correctness of \autoref{algo:one}, it is sometimes more convenient to work with the following,
obviously equivalent, definition of $D_{T'}$:
\begin{align*}
    D_{T'}=\{(\alpha_1,\dots,\alpha_k,(d_{i,j})_{1\leq i < j \leq k} ) \colon
        &\alpha_1,\dots,\alpha_k \in \sols{t}{I}, \\
        & \gamma_1,\dots,\gamma_k\in \sols{T'}{I}, \\
        & \gamma_1\cong \alpha_1,\dots,\gamma_k\cong \alpha_k, \\
        & d_{i,j} = \Delta_X(\gamma_i,\gamma_j) \text{ for } 1\leq i<j\leq k\}.
\end{align*}
The overall goal of the following lemmas is to
\begin{enumerate}
    \item show that once the status of a node $t \in V(T)$ is ``ready'', the equality $D_t = D_{T'}$ holds,
        where $T'$ is the complete subtree of $T$ rooted in $t$ (i.e., the subtree of $T$ containing
        $t$ and all of its descendants). This proves the correctness of the algorithm.
    \item provide a bound on the running time of the different steps of the algorithm.
\end{enumerate}

The first lemma, describing the size of the sets $D_t$, follows immediately from the definition
and the observation that the value of each of the entries $d_{i,j}$ is at most $|X|$.
\begin{lem}\label{lemma:sizeOfDts}
Let $Q(X)$ be an \acq, $I$ a database instance, and $\langle{}T, \lambda, r\rangle$ a join tree for $Q(X)$.
Throughout the running time of \autoref{algo:one}, for every node $t \in V(T)$ the set $D_t$ contains at most
$|R^I|^k \cdot (|X| + 1)^{\frac{k(k-1)}{2}}$ tuples, where $R^I$ is the relation in $I$ containing
the most tuples.
The size of each tuple is polynomial in the size of the input.
\end{lem}
\begin{proof}
    The polynomial size of each tuple is immediate. For the number of entries, observe that the number
    of different elements in every $\sols{t}{I}$ is $ |\sols{t}{I}| \leq |R^I|$ since $\lambda (t)$ consists of a single atom. Also, since 
    each $d_{i,j}$ describes the Hamming-Distance between two mappings with at most $|X|$ variables,
    its value is in $\{0, \dots, |X|\}$. 
    The expression $|R^I|^k \cdot (|X| + 1)^{\frac{k(k-1)}{2}}$ thus describes
    the number of all possible combinations of these values for tuples of size $k+\frac{k(k-1)}{2}$
    where the first $k$ are elements of $t(I)$ and the remaining elements from $\{0, \dots, |X|\}$.
\end{proof}

The next lemma shows  that the initialization correctly computes $D_{T'}$ for all subtrees $T'$ of
$T$ consisting of a single node, and states the running time of this step.
\begin{lem}\label{lemma:algoInitialization}
Let $Q(X)$ be an \acq, $I$ a database instance, and $\langle{}T, \lambda, r\rangle$ a join tree for $Q(X)$.
Once the ``Initialization''-step of \autoref{algo:one} is complete, the equality $D_t = D_{T'}$ holds for
all nodes $t \in V(T)$, where $T'$ is the subtree of $T$ consisting only of $t$. 
Furthermore, $D_t$ can be computed in time $\O(|\sols{t}{I}|^k \cdot k^2 \cdot |\var(t)|)$.
\end{lem}
\begin{proof}
    One helpful observation for both, correctness and the running time, is that no
    collection $\alpha_1, \dots, \alpha_k$ may occur twice, each time with different values 
    $(d_{i,j})_{1 \leq i < j \leq k}$ and $(d'_{i,j})_{1 \leq i < j \leq k}$, in $D_{T'}$:
    because the subtree $T'$ consists only of $t$, all $(\alpha_1, \dots, \alpha_k)$ are
    their only extension and thus determines the values $(d_{i,j})_{1 \leq i < j \leq k}$.
    The equality $D_t = D_{T'}$ then follows immediately from the definitions of $D_t$ and $D_{T'}$.
    
    For the time bound, observe that the given time allows one to iterate through all possible
    tuples $(\alpha_1, \dots, \alpha_k)$ with $\alpha_1, \dots, \alpha_k \in \sols{t}{I}$
    ($|\sols{t}{I}|^k$ many), and for each such tuple to compute, for each pair $\alpha_i, \alpha_j$
    with $1 \leq i < j \leq k$ (less than $k^2$ many) the value
    $d_{i,j} = \Delta_{X}(\alpha_i, \alpha_j)$ ($X \cap \var(t) \subseteq var(t)$ many variables
    to compare), which constitutes a naive implementation of the ``Initialization'' step.
\end{proof}

The next lemma will be essential in proving the ``Bottom-Up Traversal'' step of \autoref{algo:one} being
correct, and provides a bound on the running time for a single iteration of this step.
\begin{lem}\label{lemma:algoBottomUpStep}
Let $Q(X)$ be an \acq, $I$ a database instance, and $\langle{}T, \lambda, r\rangle$ a join tree for $Q(X)$.
Let $\langle{}T_1, t_1\rangle$ and $\langle{}T_2, t_2\rangle$ be two disjoint rooted subtrees
of $\langle{}T, r\rangle$ (i.e.\ $V(T_1) \cap V(T_2) = \emptyset$) such that
$t_1$ is the parent node of $t_2$ in $\langle{}T, r\rangle$, and for
$\hat{T} = T[V(T_1) \cup V(T_2)]$ consider the rooted subtree $\langle{}\hat{T}, t_1\rangle$ of
$\langle{}T,  r\rangle$. Then
\begin{align}
    D_{\hat{T}} = \{(\alpha_1, \dots, \alpha_k, (\hat{d}_{i,j})_{1\leq i < j \leq k}) \setdefsep{}
        %& \alpha_1, \dots, \alpha_k \in \sols{t_1}{I}, \nonumber\\
        %& \alpha'_1, \dots, \alpha'_k \in \sols{t_2}{I}, \nonumber\\
        & (\alpha_1, \dots, \alpha_k, (d_{i,j})_{1\leq i < j \leq k} ) \in D_{T_1}, \nonumber\\ 
        & (\alpha'_1, \dots, \alpha'_k, (d'_{i,j})_{1\leq i < j \leq k}) \in D_{T_2}, \nonumber\\
        & \alpha_i \cong \alpha'_i \text{ for } 1 \leq i \leq k, \nonumber\\
        & \hat{d}_{i,j} = d_{i,j} + d'_{i,j} - \Delta_X(\alpha_i \cap \alpha'_i, \alpha_j \cap \alpha'_j) \nonumber\\
        & \mbox{}\hspace{12em} \text{ for } 1 \leq i < j \leq k \label{equation:lemma:algoBottomUpStep}\}.
\end{align}
Also, given $D_{T_1}$ and $D_{T_2}$, the set
$D_{\hat{T}}$ can be computed in time $\O(|D|^2 \cdot k^2 \cdot |Z|)$ where 
$D$ is the larger of the two sets $D_{T_1}$ and $D_{T_2}$, and $Z$ is the larger of
the sets $\var(t_1)$ and $\var(t_2)$.
\end{lem}
\begin{proof}
 For this proof, we will use $D_{\hat{T}}$ to describe the set of tuples according to the
 initial definition, and $\hat{D}_{\hat{T}}$ for the set defined by the right hand side
 of Equation~\ref{equation:lemma:algoBottomUpStep}. We show $D_{\hat{T}} = \hat{D}_{\hat{T}}$
 by proving $D_{\hat{T}} \subseteq \hat{D}_{\hat{T}}$ and $\hat{D}_{\hat{T}} \subseteq D_{\hat{T}}$
 separately.
 
 $D_{\hat{T}} \subseteq \hat{D}_{\hat{T}}\colon$

 \noindent
 Let $e = (\alpha_1, \dots, \alpha_k, (\hat{d}_{i,j})_{1 \leq i < j \leq k})) \in D_{\hat{T}}$.
 Consider an arbitrary $(\hat{\gamma}_1, \dots, \hat{\gamma}_k) \in \ext{\hat{T}}(e)$, and define
 $\gamma_i = \hat{\gamma}_i|_{\var(T_1)}$ and $\gamma'_i = \hat{\gamma}_i|_{\var(T_2)}$ for all
 $1 \leq i \leq k$.
 By definition, $\gamma_i \in \sols{T_1}{I}$ and $\gamma'_i \in \sols{T_2}{I}$, thus we have
 \begin{align*}
     e_1 = (\gamma_1|_{\var(t_1)}, \dots, \gamma_k|_{\var(t_1)}, 
        (\Delta_X(\gamma_i, \gamma_j))_{1 \leq i < j \leq k}) &\in D_{T_1}
     \text{ and } \\
     e_2 = (\gamma'_1|_{\var(t_2)}, \dots, \gamma'_k|_{\var(t_2)}, 
        (\Delta_X(\gamma'_i, \gamma'_j))_{1 \leq i < j \leq k}) &\in D_{T_2}.
 \end{align*}
 This is the case since clearly 
 \begin{align*}
    (\gamma_1, \dots, \gamma_k) &\in 
        \ext{T_1}((\gamma_1|_{\var(t_1)}, \dots, \gamma_k|_{\var(t_1)}, (\Delta_X(\gamma_i, \gamma_j))_{1 \leq i < j \leq k})) 
    \text{ and} \\
    \revc{(\gamma'_1, \dots, \gamma'_k)} &\in 
        \ext{T_2}((\gamma'_1|_{\var(t_2)}, \dots, \gamma'_k|_{\var(t_2)}, (\Delta_X(\gamma'_i, \gamma'_j))_{1 \leq i < j \leq k})). 
\end{align*}
Now $\alpha_i = \gamma_i|_{\var(t_1)}$ and we define $\alpha'_i = \gamma'_i|_{\var(t_2)}$.
We get $\alpha_i \cong \alpha'_i$ 
for all $1 \leq i \leq k$.
Next, for $1 \leq i < j \leq k$, we have 
\begin{align*}
    \hat{d}_{i,j} &= \Delta_X(\hat{\gamma}_i, \hat{\gamma}_j) = 
\Delta_{X \cap \big(\var(T_1) \cup \var(T_2)\big)}(\hat{\gamma}_i, \hat{\gamma}_j) \\
& =
\Delta_{X \cap \big(\var(T_1) \cup 
(\var(T_2) \setminus \var(T_1))\big)}(\hat{\gamma}_i, \hat{\gamma}_j) =
\Delta_{X \cap \var(T_1)}(\hat{\gamma}_i, \hat{\gamma}_j) + 
\Delta_{X \cap \big(\var(T_2) \setminus \var(T_1)\big)}(\hat{\gamma}_i, \hat{\gamma}_j).
\end{align*}
Now
\begin{align*}
    \Delta_X(\gamma_i, \gamma_j) & = \Delta_{X \cap \var(T_1)}(\hat{\gamma}_i, \hat{\gamma}_j) \text{ and} \\
    \Delta_X(\gamma'_i, \gamma'_j) & = \Delta_{X \cap \var(T_2)}(\hat{\gamma}_i, \hat{\gamma}_j) =
        \Delta_{X \cap (\var(T_2) \setminus \var(T_1))}(\hat{\gamma}_i, \hat{\gamma}_j) + 
        \Delta_{X \cap (\var(T_2) \cap \var(T_1))}(\hat{\gamma}_i, \hat{\gamma}_j).
\end{align*}
We end up with 
$\hat{d}_{i,j} = \Delta_X(\gamma_i, \gamma_j) + \Delta_{X}(\gamma'_i, \gamma'_j) -
\Delta_{X \cap (\var(T_2) \cap \var(T_1))}(\hat{\gamma}_i, \hat{\gamma}_j)$.
Given that $\gamma_i$ and $\gamma'_i$ share exactly the variables from $\var(T_1)
\cap \var(T_2)$, we get
$\Delta_{X \cap (\var(T_2) \cap \var(T_1))}(\hat{\gamma}_i, \hat{\gamma}_j) =
\Delta_X(\gamma_i \cap \gamma'_i, \gamma_j \cap \gamma'_j)$. Because of the
connectedness condition, all variables shared between any $\gamma_i$ and $\gamma'_i$
are also contained in $\gamma_i|_{\var(t_1)}$ and $\gamma'_i|_{\var(t_2)}$ and therefore
$\Delta_X(\gamma_i \cap \gamma'_i, \gamma_j \cap \gamma'_j) = 
\Delta_X(\alpha_i \cap \alpha'_i, \alpha_j \cap \alpha'_j)$.
Thus $e \in \hat{D}_{\hat{T}}$ is verified by the tuples $e_1$ and $e_2$, which, together
with~$e$ satisfy all conditions stated on the right-hand side of
Equation~\ref{equation:lemma:algoBottomUpStep}, which concludes this direction of the
proof.

$\hat{D}_{\hat{T}} \subseteq D_{\hat{T}}\colon$
Consider an arbitrary tuple 
$\hat{e} = (\alpha_1, \dots, \alpha_k, (\hat{d}_{i,j})_{1 \leq i < j \leq k}) \in \hat{D}_{\hat{T}}$,
and let 
$e_1 = (\alpha_1, \dots, \alpha_k, (d_{i,j})_{1 \leq i < j \leq k}) \in D_{T_1}$ and
$e_2 = (\alpha'_1, \dots, \alpha'_k, (d'_{i,j})_{1 \leq i < j \leq k}) \in D_{T_2}$ 
be two tuples witnessing $\hat{e} \in \hat{D}_{\hat{T}}$ (i.e.\ $\hat{e}$, $e_1$, and $e_2$ satisfy
all conditions on the right-hand side of Equation~\ref{equation:lemma:algoBottomUpStep}).
Then $\ext{T_1}(e_1)$ and $\ext{T_2}(e_2)$ are both not empty. Choose
$(\gamma_1, \dots, \gamma_k) \in \ext{T_1}(e_1)$ and 
$(\gamma'_1, \dots, \gamma'_k) \in \ext{T_2}(e_2)$ arbitrarily. By definition,
$\Delta_X(\gamma_i, \gamma_j) = d_{i,j}$ and
$\Delta_X(\gamma'_i, \gamma'_j) = d'_{i,j}$ 
for all $1 \leq i < j \leq k$.
Because of the connectedness condition for join trees, 
\begin{enumerate}
    \item the mapping $\hat{\gamma}_i = \gamma_i \cup \gamma'_i$ is a valid mapping,
    \item $\hat{\gamma}_i|_{\var(t_1)} = \alpha_i$, and
    \item $\hat{\gamma}_i \in \sols{\hat{T}}{I}$ (and thus $\alpha_i \in \sols{t_1}{I}$). 
\end{enumerate}
Thus $(\alpha_1, \dots, \alpha_k, (\Delta_X(\hat{\gamma}_i, \hat{\gamma}_j))_{1 \leq i < j \leq k}) \in
D_{\hat{T}}$, and proving $\Delta_X(\hat{\gamma}_i, \hat{\gamma}_j) = \hat{d}_{i,j}$ concludes the proof.
Towards this goal, by an equivalent development as for proving the other direction, we get
\begin{align*}
    \Delta_X(\hat{\gamma}_i, \hat{\gamma}_j) &= \Delta_X(\gamma_i, \gamma_j) + 
        \Delta_X(\gamma'_i, \gamma'_j) - \Delta_X(\gamma_i \cap \gamma'_i, \gamma_j \cap \gamma'_j) \\
        &= d_{i,j} + d'_{i,j} - \Delta_X(\alpha_i \cap \alpha'_i, \alpha_j \cap \alpha'_j)  \\
        &= \hat{d}_{i,j}.
\end{align*}

To prove that $\hat{D}_{\hat{T}}$ can in fact be computed within the stated time bound, consider
the following naive implementation:
Iterate through all $e_1 \in D_{T_1}$, and for each such $e_1$ -- in a nested loop -- look at
each $e_2 \in D_{T_2}$ (\revc{``Loop''}). For each such pair with 
$e_1 = (\alpha_1, \dots, \alpha_k, (d_{i,j})_{1 \leq i < j \leq k})$ and
$e_2 = (\alpha'_1, \dots, \alpha'_k, (d'_{i,j})_{1 \leq i < j \leq k})$, 
check whether $\alpha_i \cong \alpha'_i$ (\revc{``Check''}).
If $\alpha_i \cong \alpha'_i$, compute $\hat{d}_{i,j}$ as defined (\revc{``Compute''}).
\revc{``Check'' requires to compare the values of $|Z|$ variables on $k$ pairs $(\alpha_i, \alpha'_i)$
(possible in $\O(k \cdot |Z|)$ time),
and ``Compute'' computes $\frac{k(k-1)}{2}$ many values (thus in $\O(k^2)$). These two steps
are performed $\O(|D_{T_1}| \cdot |D_{T_2}|)$ times (number of iterations of ``Loop''). 
Despite ``Check'' and ``Compute'' being sequential, for simplicity we bound the running time
by $\O(|D|^2 \cdot k^2 \cdot |Z|)$ instead of $\O(|D|^2 \cdot (k^2 + k \cdot |Z|))$.}
\end{proof}

With this result at hand, we can show that the bottom-up traversal of the join tree is correct.
\begin{lem}\label{lemma:algoBottomUp}
Let $Q(X)$ be an \acq, $I$ a database instance, and $\langle{}T, \lambda, t\rangle$ a join tree for $Q(X)$.
At the end of every iteration, the ``Bottom-Up Traversal'' step of \autoref{algo:one} guarantees the
following two properties:
\begin{enumerate}
    \item For all nodes $t \in V(T)$ with status ``ready'', the equality $D_t = D_{T'}$ holds, where
        $T'$ is the subtree of $T$ consisting of $t$ and all its descendants.
    \item For a node $t \in V(T)$, let $t_1, \dots, t_p$ be the child nodes with status ``processed''.
        Then $D_t = D_{T'}$ where $T'$ is the subtree of $T$ consisting of $t$ and all $t_i$ and
        all their descendants, for $1 \leq i \leq p$.
\end{enumerate}
\end{lem}
\begin{proof}
    We show both properties by induction on the number of steps in the bottom-up traversal of 
    the join tree. 
    Throughout this proof, for a node $t \in V(T)$, we will use $T_t$ to denote the complete 
    subtree of $T$ rooted in $t$, i.e.\ the subtree of $T$ containing $t$ and all its descendants.
    
    For the base case, consider the situation before the first iteration of the bottom-up
    traversal. At this point, the set of nodes with status ``ready'' are exactly the leaf nodes.
    Since they have no child nodes, the statement $D_t = D_{T'}$ in property (1) is equivalent to 
    $D_t = D_{T[\{t\}]}$.
    By \autoref{lemma:algoInitialization}, this equality holds for all nodes once the ``Initialization'' 
    step is finished. 
    We next observe that there are no nodes with status ``processed''. Thus for all nodes $t \in V(T)$
    property (2) also states $D_t = D_{T[\{t\}]}$, which again holds because of 
    \autoref{lemma:algoInitialization}.

    For the induction step, consider the node $t \in V(T)$ for which $D_t$ was updated to $\revc{\Dnew_t}$ in the 
    ``Bottom-Up Traversal'' step, and let $t'$ be the child node of $t$ that was used to compute 
    the update.
    As induction hypothesis, we know that the first property holds for the child node $t'$
    (status before the step: ``ready''), and the second property holds for $t$ w.r.t.\ all the 
    child nodes $t_1, \dots, t_{p}$ of $t$ with status ``processed'' (possibly none).
    We have to show that after the ``Bottom-Up Traversal'' step:
    \begin{enumerate}[a)]
        \item The second property holds for $t$ and the child nodes $t_1, \dots, t_p, t'$.
        \item If $t'$ was the only remaining child node of $t$ with a status different from ``processed'',
            then the first property now holds for $t$.
    \end{enumerate}
    To prove a), let $T' = T[\{t\} \cup V(T_{t_1}) \cup \dots V(T_{t_p})]$ and
    $\hat{T} = T[\{t\} \cup V(T_{t_1}) \cup \dots \cup V(T_{t_p}) \cup V(T_{t'}) ]$. 
    The induction hypothesis guarantees that \autoref{lemma:algoBottomUpStep} applies (i.e.\ all
    the preconditions are satisfied w.r.t.\ $T'$, $T_{t'}$, and $\hat{T}$). Observe that the set
    described in \autoref{lemma:algoBottomUpStep} is exactly the set $\revc{\Dnew_t}$ computed from
    $D_t$ by the ``Bottom-Up Traversal'' step. We thus have $D_t = D_{T'}$ by the induction hypothesis, 
    and $\revc{\Dnew_t} = D_{\hat{T}}$ by \autoref{lemma:algoBottomUpStep}, which completes the proof of a).
    
    For b), observe that the only node whose status can switch to ``ready'' is $t$. If this happened
    at the end of the step, then $t'$ was the last child of $t$ whose status was not ``processed'',
    and we now have $\hat{T} = T_t$. Thus $D_t = D_{T_t}$ follows immediately from a), concluding
    the proof of the lemma.     
\end{proof}
\noindent
We now have everything in place to prove \autoref{theorem:acqXP}.
\begin{proof}[Proof (of \autoref{theorem:acqXP})]
    We start by proving the correctness of the algorithm, before discussing its running time.

    For the correctness, from \autoref{lemma:algoBottomUp} we know that once the
    ``Bottom-Up Traversal'' step is finished (i.e.\ there is no more node with status
    ``not-ready''; in other words, the root has status ``ready'' and all other nodes
    have status ``processed''), then $D_r = D_T$ ($r$ is the root of $T$). As a result, for any 
    $e = (\alpha_1, \dots, \alpha_k, (d_{i,j})_{1 \leq i < j \leq k}) \in D_r$
    and every $(\gamma_1, \dots, \gamma_k) \in \ext{T}(e)$ we have 
    $\gamma_i \in \sols{T}{I}$ and 
    $\gamma_i|_X \in Q(I)$ for all $1 \leq i \leq k$.
    Hence
    $\delta(\gamma_1, \dots, \gamma_k) = 
    f((\Delta_X(\gamma_i, \gamma_j))_{1 \leq i < j \leq k}) =
    f((d_{i,j})_{1 \leq i < j \leq k})$
    (with $f$ being the polynomial time computable function defining $\delta$).
    
    Thus the correctness of the ``Finalization'' step follows immediately from
    \begin{align*}
        \max_{\substack{\gamma_1, \dots, \gamma_k \in \sols{Q}{I}\\ 
            \gamma_i \neq \gamma_j \text{ for } i \neq j}} \delta(\gamma_1, \dots, \gamma_k) &= 
        \max_{\substack{\gamma_1, \dots, \gamma_k \in \sols{T}{I}\\ 
            \gamma_i|_X \neq \gamma_j|_X \text{ for } i \neq j}} \delta(\gamma_1|_X, \dots, \gamma_k|_X) \\
        &= \max_{\substack{\gamma_1, \dots, \gamma_k \in \sols{T}{I} \\
                \Delta_X(\gamma_i, \gamma_j) > 0 \text{ for } i \neq j}}
                    f((\Delta_X(\gamma_i, \gamma_j))_{1 \leq i < j \leq k}) \\
        &= \max_{\substack{(\alpha_1, \dots, \alpha_k, (d_{i,j})_{1 \leq i < j \leq k}) \in D_r \\
            d_{i,j} > 0 \text{ for } 1 \leq i < j \leq k}}
                    f((d_{i,j})_{1 \leq i < j \leq k}).
    \end{align*}
    
    For the bound on the running time, by \autoref{lemma:algoInitialization}, the 
    ``Initialization'' step takes time in 
    $\O(|\sols{t}{I}|^k \cdot k^2 \cdot |\var(t)|)$ for each
    node, i.e.\ $\O(|R^I|^k \cdot k^2 \cdot |\var(A)| \cdot |Q|)$ in total (the join tree
    contains one node for each atom in $Q(X)$), where $R^I$ is the relation in $I$ with the highest
    number of tuples, and $A$ is the atom in $Q$ with the highest number of variables.
    
    By \autoref{lemma:algoBottomUpStep}, one iteration of the ``Bottom-Up Traversal'' step
    takes time in $\O((|R^I|^k \cdot (|X| + 1)^{\frac{k(k-1)}{2}})^2 \cdot k^2 \cdot |\var(A)|)$
    using the size bound on $D_t$ from \autoref{lemma:sizeOfDts}.
    Since every node (except the root node) is merged into its parent node exactly once, we get 
    $\O(|R^I|^{2k} \cdot (|X| + 1)^{k(k-1)} \cdot k^2 \cdot |\var(A)|) \cdot |Q|)$ in total.
    The ``Finalization`` step takes time $\O(|R^I|^{2k} \cdot (|X| + 1)^{k(k-1)} \cdot \pol_f(|X|,k))$,
    where $\pol_f(|X|,k)$ is a polynomial describing the time needed to compute the function 
    $f((d_{i,j})_{1 \leq i < j \leq k})$. Thus the running time of the ``Bottom-Up Traversal''
    dominates the running time of the ``Initialization'' step, which is why we can omit it,
    providing the running time stated in the theorem.
\end{proof}

\autoref{theorem:acqXP} shows that the algorithm \emph{decides} in $\xp$ the existence
of a diversity set with a given diversity.
Computing a witness diversity set now means computing one
element $(\gamma_1, \dots, \gamma_k) \in \ext{T}(e)$ for some $e \in D_T$ with 
$f( (d_{i,j})_{1\leq i < j \leq k}  ) \geq d$
and $d_{i,j} \neq 0$ for all $i,j$.
Similarly to the construction of an answer tuple by the Yannakakis algorithm for \cqs,
we can compute an arbitrary element from $\ext{T}(e)$ by making use of the information stored
in the final sets $\rho_{D_{t}}(e)$.
By construction, for every node $t \in V(T)$ and every entry $e \in D_{T'}$,
the final set $\rho_{D_t}(e)$ contains exactly one pair $(t', e')$ for every child node $t'$ of $t$.
Moreover, for the mappings $\alpha_1, \dots, \alpha_k$ from $e$ and $\alpha'_1, \dots, \alpha'_k$
from $e'$, $\alpha_i \cong \alpha'_i$ holds for all $1 \leq i \leq k$, hence $\alpha_i \cup \alpha'_i$
are again mappings.
Thus, to compute the desired witness $(\gamma_1, \dots, \gamma_k) \in \ext{T}(e)$ for the chosen $e \in D_T$,
start with $(\alpha_1, \dots, \alpha_k)$ from $e$, take all $(t', e')$ from 
$\rho_{D_r}(e)$, extend each $\alpha_i$ with $\alpha'_i$ from $e'$, and repeat this step recursively.

\begin{figure}[htp]
\begin{center}
\includegraphics[width=\textwidth]{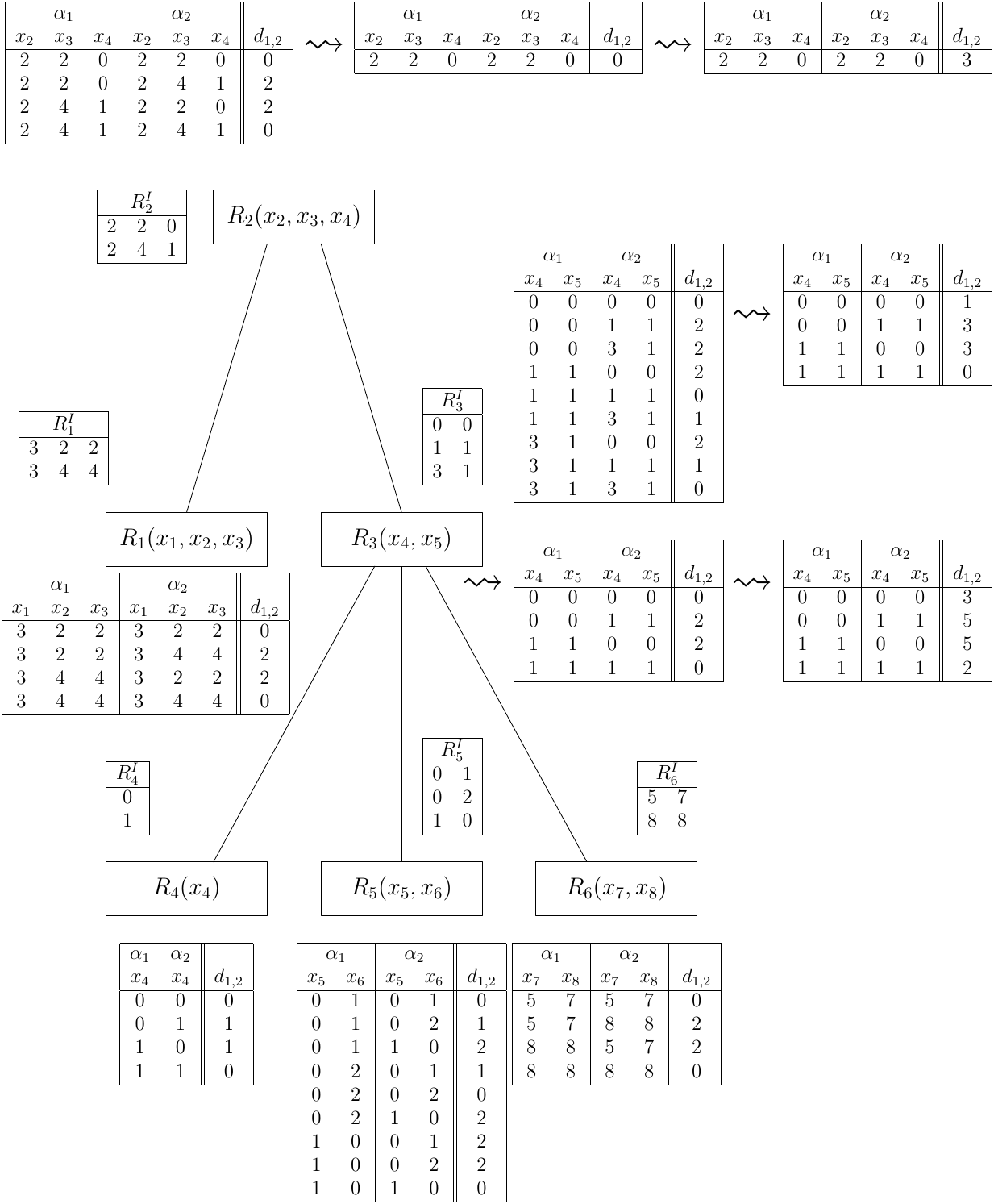}
\caption{Example Execution of the Basic Algorithm.}
\label{fig:acqExecution}
\end{center}
\end{figure}

\begin{exa}
\revc{
An example execution of the basic algorithm for $k=2$ on the query 
\[Q(x_1,\dots, x_8):-R_1(x_1,x_2,x_3)\land R_2(x_2,x_3,x_4) \land R_3(x_4,x_5) \land R_4(x_4) \land R_5(x_5,x_6) \land R_6(x_7,x_8)\] 
which, together with a possible join tree, is shown in 
Figure \ref{fig:acqExecution}.
The database $I$ 
consists of (very small) relations 
$R^I_1, \dots, R^I_6$ and each relation $R^I_i$
is shown in the figure next to the node $R_i$.
For a node $t$, the set $D_{t}$ is computed by considering the subtrees 
rooted at the children of $t$ from left to right.
For the sake of succinctness, tuples $(\alpha_1,\alpha_2,d_{1,2})\in D_{t}$ are omitted if there is a strictly better tuple, i.e., a $(\alpha_1,\alpha_2,d_{1,2}') \in D_{t}$ such that $d_{1,2}<d_{1,2}'$.
(Formally, this is only justified if the aggregator is monotone.)
We do not specify an aggregator and skip the ``Finalization'' step as we are only looking for a diverse pair.
This pair is $\{(3,2,2,0,0,1,5,7), (3,2,2,0,0,2,8,8)\}$.
}

\revc{We now discuss the ``Initialization'' and ``Bottom-Up Traversal'' of 
\autoref{algo:one} in more detail: 
We first carry out the 
initialization step of \autoref{algo:one} for all nodes. That is, 
we set up a table with all possible pairs of tuples from $R^I_i$ 
together with their Hamming distance (recall that we are looking for pairs since we have  
$k = 2$ in this example). 
% The diversity of each pair corresponds to their Hamming distance. 
In particular, if a pair consists of two identical tuples, then we get
a distance of 0.
}

\revc{
We next discuss the result of carrying out the bottom-up traversal.
To this end, we inspect the two internal nodes $R_3$ and $R_2$.
First, look at the tables to the right of the node $R_3$: the tuple $(3,1)$ has no join partner in the leftmost
child (= $R_4$). Hence, in the table to the right of the initial one, 
we delete all pairs that contain the tuple $(3,1)$. Therefore, we only consider pairs built from the first two tuples 
in $R^I_3$, i.e., 
$(0,0)$ and $(1,1)$. Clearly, extending these pairs to the leftmost
child does not add to the distance, since that node (= $R_4$) has no additional variable. 
In the left table below, we show the result of extending these 4 pairs of tuples to the second child.
It turns out that, for the first 3 pairs, the maximum achievable distance increases by 1 because we could 
extend the tuples of such a pair in two different ways to $x_6$. Now let us also look at the last pair in this 
table, i.e., combining $(1,1)$ with $(1,1)$. That is, in both tuples, $x_5$ is set to $1$. But when we look at 
the table corresponding to $R_5$, it turns out that the only possible extension to $x_6$ is $1$. Hence, the 
distance of this pair cannot be increased by an extension to $R_5$ and it remains 2.
We then carry out the bottom-up step also from the child node $R_6$ to $R_3$. Now we can indeed extend
the tuples of each pair to different values of $x_7$ and also $x_8$, which leads to an increase of the distance
by 2. That is, we end up with (maximally achievable) distances $3,5,5,$ and $2$, respectively, for the node $R_3$.  
}

\revc{
We finally also discuss the tables above the root node. The leftmost table is the result of the initialization step. 
For the bottom-up step from the left child (= $R_1$) to the root node, we observe that only the tuple
$(2,2,0)$ of $R^I_2$ has a join partner in $R^I_1$. Moreover, if we fix $x_2=2$ and $x_3=2$, then there exists only 
one possible extension to $x_1$ in $R^I_1$, namely $x_1 = 3$. Hence, the second table above node $R_2$ consists
of a single pair and its initial distance (namely 0) cannot be increased by an extension to the left child. 
Now consider also the right child of $R_2$. We are only considering the pair from  $R^I_2$ where both tuples are
the same, namely $(2,2,0)$. We see in the last table attached to $R_3$ that the maximum distance achievable by
pairs where both tuples have $x_4 = 0$ is 3. Hence, this is then also the maximum distance achievable by 
the only pair in the last table attached to $R_2$. By tracing back top-down the pairs from the bottom-up traversal 
which contributed to the maximally achievable distance at the parent node, we get the pair $(3,3,2,0,0,1,5,7)$ and
$(3,3,2,0,0,2,8,8)$ of query answers with maximum distance. 
\hfill $\Diamond$
}
\end{exa}

\subsubsection{W[1]-Hardness} 
\label{sec:wonehardness}

Having proved $\xp$-membership \revc{in} combined complexity of the \DiverseACQ problem
in \autoref{theorem:acqXP}, we now show that, for any ws-monotone diversity measure, a stronger result in the form of $\fpt$-membership is
very unlikely to exist. 
More specifically, we prove $\wone$-hardness for combined complexity in these cases.
\revc{The reduction we use only takes polynomial time and, thus, we spontaneously prove $\np$-hardness of \DiverseACQstrmon in combined complexity when considering the problem unparameterized}

\begin{thm} \label{theorem:acqCCHardness}
    The problem \DiverseACQstrmon, parameterized by the
    size~$k$ of the diversity set, is $\wone$-hard \revc{in combined complexity}.
    It remains $\wone$-hard even if 
    all relation symbols are of arity at most two and 
    $Q(X)$ contains no existential variables.    
    \revc{Furthermore, viewed as an unparameterized problem, \DiverseACQstrmon is $\np$-hard in combined complexity}
\end{thm} 
\begin{proof}
We reduce from the \IndependentSet problem parameterized by the size of the independent set.

Let $(G, s)$ be an arbitrary instance of \IndependentSet with
$V(G) = \{v_1, \dots, v_n\}$ and
$E(G) = \{e_1, \dots, e_m\}$.
We define an instance $\langle I, Q, k, d \rangle$ of \DiverseACQstrmon as follows. 
The schema consists of a  relation symbol
$R$ of arity one
and $m$ relation symbols $R_1, \dots, R_m$ of arity two.
The \cq $Q(X)$ is defined as 
\[
Q(v, x_1, \dots, x_m) := R(v) \land R_1(v, x_1) \land \dots \land R_m(v, x_m)
\]
and the database instance $I$ with $\dom(I) = \{0, 1, \dots, n\}$ is
\begin{align*}
    R^I =& \{ (i) \setdefsep v_i \in V(G)\} \mbox{ and}\\
    R_j^I =& \{ (i, i) \setdefsep v_i \text{ is not incident to } e_j\} \cup 
        \{ (i, 0) \setdefsep v_i \text{ is incident to } e_j\}
            \text{ for all } j \in \{1, \dots, m\}.
\end{align*}
Finally, set $k = s$ and $d = f(m+1, \dots, m+1)$,
where $f$ is the aggregator of \revc{$\deltastrmon$ aggregating the value $m+1$ exactly $\binom{k}{2}$ times}.
Clearly, this reduction is feasible in polynomial time and the resulting problem instances 
satisfy all the 
restrictions stated in the theorem.
The correctness of this reduction depends on two main observations.
\begin{enumerate}[O1)]
    \item For each $i \in \{1, \dots, n\}$, independently of $G$, there exists exactly
        one solution $\gamma_i \in Q(I)$ with $\gamma_i(v) = i$, and these are in fact the only
        solutions in $Q(I)$.
        Thus, there is a natural one-to-one association between vertices $v_i\in V(G)$ and solutions 
        $\gamma_i\in Q(I)$.
    \item Due to ws-monotonicity, the desired diversity $d =  f(m+1, \dots, m+1)$  can only be achieved by $k$ solutions that 
        pairwisely differ  on all variables.
\end{enumerate}

\noindent
Observation O1 is immediate: 
$R(\gamma(v)) \in R^I$ if and only if $\gamma(v) \in \{1, \dots, n\}$, and for every
$i \in \{1, \dots, n\}$ and $j \in \{1, \dots, m\}$, there exists exactly one pair
$(i, b) \in R_j^I$ (with $b$ being either $0$ or $i$).
For Observation O2, note that the Hamming distance between two answers is at most $m+1$.
Thus, if two answers $\gamma_i,\gamma_j$ are equal on some variable, $\Delta(\gamma_i,\gamma_j) < m + 1$, and hence due to ws-monotonicity, $\deltastrmon(\gamma_1,\dots, \gamma_k) = f((\Delta(\gamma_i,\gamma_j))_{1\leq i < j \leq k}) < f(m+1,\dots, m+1)=d$.

Observation O2 allows us to prove the correctness of the reduction by showing that
$G$ has an independent set of size $s$ if and only if there exists a diversity set
of size $k$ where all answers differ pairwise on all variables.

To do so, first
assume that there exists an independent set $S \subseteq V(G)$ of size $s$ in $G$.
We define the diversity set as $D = \{\gamma_i \setdefsep v_i \in S\}$.
By Observation O1, $D$ is well-defined and thus contains $k$ answers.
We show that for any two distinct solutions $\gamma_i, \gamma_j \in D$ we
have $\gamma_i(x) \neq \gamma_j(x)$ for all variables $x \in \var(Q)$. 
To do so, first note that for all solutions $\gamma \in Q(I)$ and all variables $x \in \var(Q)$,
the fact that $\gamma(x) \neq 0$ implies $\gamma(x) = \gamma(v)$.
Because of Observation O1, this implies that $\gamma_i(x) = \gamma_j(x)$ is only possible if 
$\gamma_i(x) = \gamma_j(x) = 0$ (since O1 implies that there do not exist any two distinct
solutions $\gamma_i, \gamma_j \in Q(I)$ with $\gamma_i(v) = \gamma_j(v)$).
Thus, towards a contradiction, assume $\gamma_i(x_\ell) = \gamma_j(x_\ell) = 0$ for
some $\ell \in \{1, \dots, m\}$.
Then both, $(i, 0)$ and $(j, 0)$ must be contained in $R_\ell$, and by the definition
of $I$ this implies that both, $v_i$ and $v_j$ are incident to $e_\ell$.
This however contradicts that $v_i$ and $v_j$ are both part of the same
independent set, which proves that any two solutions $\gamma_i \neq \gamma_j$ must
differ on all variables.

Next assume that there exists a diversity set $D \subseteq Q(I)$ of size $k$
such that all distinct answers in $D$ differ on all variables. 
We define a set $S = \{v_i \setdefsep \gamma_i \in D\}$.
By Observation O1, this set is well-defined and contains exactly
$k$ vertices. Towards a contradiction, assume that $S$ contains two
adjacent vertices $v_i \neq v_j$ and let $e_\ell$ be the edge connecting
$v_i$ and $v_j$. By definition of $I$, we get $(i, 0) \in R^I$ and 
$(j, 0) \in R^I$. However, this implies
$\gamma_i(x_\ell) = 0 = \gamma_j(x_\ell)$.
This, however, contradicts the assumption that all solutions differ
pairwise on all variables, which concludes the proof.
\end{proof}

\subsubsection{Speeding up the Basic Algorithm} \label{sec:speedups}
\autoref{algo:one} works for any polynomial-time computable diversity measures $\delta$. 
To compute
the diversity at the root node, we needed to distinguish between all the possible values
for $d_{i,j}$ ($1 \leq i < j \leq k$), which heavily increases the size of the sets $D_t$.
The reason we had to explicitly distinguish all 
these values
in the basic algorithm
is that, in general,
given two collections $(\gamma_1, \dots, \gamma_k)$ and $(\gamma'_1, \dots, \gamma'_k)$ of mappings 
that agree on the shared variables,  we cannot derive 
$\delta(\hat{\gamma_1}, \dots, \hat{\gamma}_k)$
for $\hat{\gamma}_i = \gamma_i \cup \gamma'_i$
from 
$\delta(\gamma_1, \dots, \gamma_k)$ and 
$\delta(\gamma'_1, \dots, \gamma'_k)$.
However, for specific diversity measures, this is possible. As a result, significantly less 
information needs to be maintained, as will now be exemplified for $\deltasum$.

\begin{thm} \label{theorem:DiverseACQsum}
    The \DiverseACQsum problem is in $\fpt$ \revc{in} query complexity 
    when parameterized by the size $k$ of the diversity set.
    More specifically,  \DiverseACQsum for an ACQ $Q(X)$, a database instance $I$, and integers $k$ and $d$,
    can be solved in time 
    $\O(|R^I|^{2k} \cdot 2^{k(k-1)} \cdot \pol(|Q|, k))$, where~$R^I$ is the relation from $I$ with the most tuples and 
    $\pol(|Q|, k)$ is a polynomial in $|Q|$ and $k$.
\end{thm}

\begin{proof}
Note that $\pol(|Q|, k)$ is the same as in \autoref{theorem:acqXP}.
For query complexity, the size~$|R^I|$ of a relation in $I$ is considered as constant. Hence, 
the above-stated upper bound on the asymptotic complexity indeed entails $\fpt$-membership.
To prove this upper bound, the crucial property  is that for 
a collection of mappings $\gamma_1, \dots, \gamma_k$ over variables $Z$, the equality
$\deltasum(\gamma_1, \dots, \gamma_k) = \sum_{z \in Z}\deltasum(\gamma_1|_{z}, \dots, \gamma_k|_{z})$
holds.  
Hence, in principle, it suffices to store in~$D_{T'}$ for each collection 
$(\alpha_1, \dots, \alpha_k)$ with $\alpha_i \in \sols{t}{I}$ ($t$ being the root of $T'$)
such that there exists $\gamma_i \in \sols{T'}{I}$ with $\gamma_i \cong \alpha_i$ 
(for all $1 \leq i \leq k$) the value 
\[d_{T'}(\alpha_1, \dots, \alpha_k) = 
\max_{\substack{\gamma_1, \dots, \gamma_k \in \sols{T'}{I}\\\text{s.t. } 
\gamma_i \cong \alpha_i \text{ for all } i}} \deltasum(\gamma_1|_X, \dots, \gamma_k|_X).\]
I.e., each entry in $D_{T'}$ now is of the form $(\alpha_1, \dots, \alpha_k, v)$ with
$v = d_{T'}(\alpha_1, \dots, \alpha_k)$. 
In the bottom-up traversal step of the algorithm, when updating some $D_t$ to $\revc{\Dnew_t}$ by merging
$D_{t'}$, for every entry $(\alpha_1, \dots, \alpha_k, v) \in D_t$ there exists an entry
$(\alpha_1, \dots, \alpha_k, \bar{v}) \in \revc{\Dnew_t}$ if and only if there exists at least one 
$(\alpha'_1, \dots, \alpha'_k, v') \in D_{t'}$ such that $\alpha_i \cong \alpha'_i$ for
$1 \leq i \leq k$. Then $\bar{v}$ is
\[
    \bar{v} = \max_{\substack{(\alpha'_1, \dots, \alpha'_k, v') \in D_{t'}\\ \text{ s.t. }
        \alpha_i \cong \alpha'_i \text{ for all } i}} 
        (v +  v' - \deltasum((\alpha_1\cap \alpha'_1)|_{X}, \dots, (\alpha_k\cap \alpha'_k)|_{X})).
\]

In order to make sure that 
the answer tuples in the final diversity set are pairwise distinct, 
the following additional information must
be maintained at each $D_{T'}$: from the partial solutions $\alpha_1, \dots, \alpha_k$ it
is not possible to determine whether the set of extensions $\gamma_1, \dots, \gamma_k$ contains
duplicates or not. Thus, similar to the original values $d_{i,j}$ describing the pairwise
diversity of partial solutions, we now include binary values $b_{i,j}$ for $1 \leq i < j \leq k$
that indicate whether extensions $\gamma_i$ and $\gamma_j$ of $\alpha_i$ and $\alpha_j$ to
$\var(T')$ differ on at least one variable of $X$ ($b_{i,j} = 1$) or not in order to be part of
$\ext{T'}(e)$.
This increases the maximal size of $D_{T'}$ to $|R^I|^{2k} \cdot 2^{k(k-1)}$. The bottom-up
traversal step can be easily adapted to consider in the computation of $\bar{v}$ for an entry
in $\revc{\Dnew_{t}}$ only those entries from $D_t$ and $D_{t'}$ that are consistent with the values of
$b_{i,j}$, giving the stated running time.
\end{proof}

Actually, 
if we drop the condition that the answer tuples in the final diversity set must be 
pairwise distinct, the query complexity of 
\DiverseACQsum can be further reduced. 
Clearly, in this case, we can drop the binary values $b_{i,j}$ for $1 \leq i < j \leq k$ from the entries in 
$D_{T'}$, which results in a reduction of the asymptotic complexity to $\O(|R^I|^{2k} \cdot \pol(|Q|, k))$. 
At first glance, this does not seem to improve on the $\fpt$-membership result. 
However, a further, generally applicable improvement (not restricted to a particular aggregate function and 
not restricted to query complexity) is possible via the observation that the
basic algorithm computes (and manages) redundant information: for
an arbitrary node $t \in V(T)$ and set $D_{t}$, 
if $D_t$ contains an entry of the form 
$(\alpha_1, \dots, \alpha_k, \dots)$,
then $D_t$ also 
contains entries of the form 
$(\alpha_{\pi(1)}, \dots, \alpha_{\pi(k)}, \dots)$
for all permutations $\pi$ of $(1, \dots, k)$.
But we are ultimately interested in {\em sets} of answer tuples and do not distinguish 
between permutations of the members inside a set.
Keeping these redundant entries made the algorithm conceptually simpler and had no significant
impact on the running times (especially since we assume $k$ to be small compared to the size of the
relations in $I$). 
However, given the improvements for \DiverseACQsum from \autoref{theorem:DiverseACQsum} and 
dropping the binary values $b_{i,j}$ for $1 \leq i < j \leq k$ from the entries in $D_t$,
we can get a significantly better complexity classification:

\begin{thm} \label{theorem:querycomplexitypolytime}
    The problem \DiverseACQsum is in $\ptime$ \revc{in} query complexity when the diversity set may 
    contain duplicates and $k$ is given in unary.
\end{thm}
\begin{proof}
\revc{We claim} the number of rows in $D_t$ for any $t \in V(T)$ to be in $\O(k^{|\sols{t}{I}|-1})$.
In the following, we verify the claim. 

To remove redundant rows from the sets $D_t$, we introduce some order $\preceq$ on partial solutions
$\alpha \in \sols{t}{I}$ for each $t \in V(T)$ (e.g.\ based on some order on the domain elements),
and only consider such collections
$\alpha_1, \dots, \alpha_k \in \sols{t}{I}$ where $\alpha_1 \preceq \dots \preceq \alpha_k$ together
with the value $d_{T'}(\alpha_1, \dots, \alpha_k)$.
Thus the number of such different collections is described by
$
\binom{|\sols{t}{I}|+k-1}{k}
$.
Applying basic combinatorics we get
\[
\binom{|\sols{t}{I}|+k-1}{k} = 
\binom{|\sols{t}{I}|+k-1}{\revc{(|\sols{t}{I}|+k-1) - k} } = 
\binom{|\sols{t}{I}|+k-1}{|\sols{t}{I}| - 1}.
\]
By definition, this is the same as
\[
\frac{(|\sols{t}{I}|+k-1) \cdot (|\sols{t}{I}|+k-2) \cdot \ldots \cdot (k + 1)}{(|\sols{t}{I}| - 1)!}
\leq 
(|\sols{t}{I}|+k)^{|\sols{t}{I}|-1}.
\]
Since we assume query complexity, we consider the size of $I$ to be a constant. 
Thus, since $\lambda(t)$ consists of a single atom,
also $|\sols{t}{I}|$ can be considered to be a constant. As a result we have that
$(|\sols{t}{I}|+k)^{|\sols{t}{I}|-1}$ is in $\O(k^{|\sols{t}{I}|-1})$ as claimed.
\end{proof}

\subsection{Data Complexity}
\label{sec:CQ-data}

We now inspect the data complexity of \DiverseACQ both from the parameterized and non-parameterized point of view. 
For the parameterized case, we will improve the $\xp$-membership result from \autoref{theorem:acqXP}
(for combined complexity) to $\fpt$-membership for arbitrary monotone aggregate functions. 
Actually, by considering the query as fixed, 
we now allow arbitrary FO-queries, whose evaluation is well-known to be 
feasible in polynomial time (data complexity) \cite{DBLP:conf/stoc/Vardi82}.
Thus, as a preprocessing step, we can evaluate $Q$ and store the result in a table $R^{I}$.
We may therefore assume w.l.o.g.\ that the query is of the form 
$Q(x_1,\dots,x_m) :=  R(x_1,\dots,x_m)$
and the database $I$ consists of a single relation $R^I$. 
 
To show $\fpt$-membership, 
we apply a problem reduction that allows us to iteratively reduce the size of the 
database instance until it is bounded by a function of $m$ and $k$, i.e., the query and the parameter.
Let $X=\{x_1,\dots,x_m\}$ and define
$\binom{X}{s}:= \{Z \subseteq X : |Z| = s\}$
for $s \in \{0, \dots, m\}$.
Moreover, for every assignment $\alpha\colon Z \rightarrow \dom(I)$ with $Z \subseteq X$ 
let $Q(I)_{\alpha}:=\{\gamma\in Q(I)\colon \gamma \cong \alpha\}$, i.e., the set of 
answer tuples that coincide with $\alpha$ on $Z$.
The key to our problem reduction
is applying the following reduction rule $\textbf{Red}_t$ 
for $t \in \{1, \dots, m\}$ exhaustively in order $\textbf{Red}_1$ through $\textbf{Red}_m$:

\medskip
\noindent
$(\textbf{Red}_t)$ If for some $\alpha\colon Z \rightarrow \dom(I)$ 
with $Z\in \binom{X}{m- t}$,
the set $Q(I)_{\alpha}$ has at least $t!^{2}\cdot k^t$ elements, then do the following:
    select (arbitrarily) $t\cdot k$ solutions $\Gamma \subseteq Q(I)_{\alpha}$ that 
    pairwisely differ on all variables in $X\setminus Z$. 
    Then remove the tuples corresponding to assignments $Q(I)_{\alpha} \setminus \Gamma$ from~$R^I$.

\smallskip

The intuition of the reduction rule is best seen by looking at 
$\textbf{Red}_1$. Our ultimate goal is to achieve maximum diversity by selecting $k$ answer
tuples. 
Now suppose that we fix all but 1 position -- say $x_1$ -- in the answer relation
$R^I$ \revc{
to be equal to some assignment $\alpha$.
Furthermore, let $Q(I)_\alpha \subseteq R^I$ be the matching tuples and 
$\Gamma \subseteq Q(I)_{\alpha}$ be $k$-many of these matching tuples, chosen arbitrarily.
Now, given a diversity set $D\subseteq Q(I)$, we claim that we can replace every element $\gamma \in D \cap (Q(I)_{\alpha} \setminus \Gamma)$ with an element in $\Gamma$ while preserving optimality.
This means that it is safe to remove $Q(I)_{\alpha} \setminus \Gamma$ from $R^I$.
The claim holds as $\gamma$ and $\Gamma$ agree on all positions but $x$.
Thus, we only need to find a $\gamma'\in \Gamma$ that differs from each element in 
$D\setminus \{\gamma\}$ on $x$ as then $\gamma'$ is at least as far away from those elements as $\gamma$ is.
Such an element always exists due to $\Gamma$ containing more elements than $D\setminus \{\gamma\}$ has unique $x$-values.
}

This can be generalized to fixing fewer positions but the intuition stays the same. 
When fixing $m-t$ positions, there is also no need to retain all different value combinations in the remaining $t$ positions.
Concretely, if there exist at least $t!^2\cdot k^{t}$ different value combinations (possibly sharing values on some positions), there also exist $t \cdot k$ tuples with pairwise maximum Hamming distance on the remaining $t$ positions (no shared values pairwise) and it is sufficient to only keep those.
\revc{Note that here a recursive argument is needed to ensure the existence of the $t \cdot k$ pairwise maximally distant tuples and, hence, it is necessary to first apply $\textbf{Red}_{t-1}$ exhaustively before we can continue with $\textbf{Red}_{t}$.}

Formally, the crucial properties of the reduction rule $\textbf{Red}_t$ with 
$t \in \{1, \dots, m\}$ is as follows: 
\begin{lem}\label{lem:claimA}
Let $Q$ be a \cq of the form $Q(x_1, \dots , x_m) := R(x_1, \dots, x_m)$, $I$ a corresponding 
database, $t \in \{1, \dots, m \}$ and suppose that all sets $Q(I)_{\alpha'}$ with 
$\alpha' \colon Z' \rightarrow \dom(I)$
and $Z'\in \binom{X}{m-(t-1)}$ have cardinality at most $(t-1)!^2\cdot k^{t-1}$. 
Then the reduction rule $\textbf{Red}_t$ is well-defined and safe. That is: 
\begin{itemize}
    \item {\em ``well-defined''.} If for some $\alpha: Z \rightarrow \dom(I)$
    with $Z\in \binom{X}{m- t}$, 
    the set $Q(I)_{\alpha}$ has at least $t!^{2}\cdot k^t$ elements, then there exist at least $t\cdot k$ solutions $\Gamma \subseteq Q(I)_{\alpha}$ that pairwisely differ on all variables in $X\setminus Z$.
    \item {\em ``safe''.} Let $\Iold$ denote the database instance before 
    an application of $\textbf{Red}_t$ and let $\Inew$ denote its state after applying  $\textbf{Red}_t$. Let $\gamma_1, \dots, \gamma_k$ be pairwise distinct solutions in $Q(\Iold)$. Then 
    there exist pairwise distinct solutions $\gamma'_1, \dots, \gamma'_k$ in $Q(\Inew)$ with 
    $\delta(\gamma'_1, \dots, \gamma'_k) \geq \delta(\gamma_1, \dots, \gamma_k)$, i.e., 
    the diversity achievable before deleting tuples from the database can still be achieved after
    the deletion.
\end{itemize}
Moreover, a set of $t\cdot k$ solutions $\Gamma \subseteq Q(I)_{\alpha}$ that pairwisely differ on all variables in $X\setminus Z$ 
can be computed by iteratively choosing solutions $\gamma_i$ (for $i \in \{1, \dots, t \cdot k\}$) arbitrarily from $Q(I)_{\alpha}$ 
that differ from all solutions $\gamma_1, \dots, \gamma_{i-1}$ on all variables in $X\setminus Z$.
\end{lem}
\begin{proof}
Let $t \in \{1, \dots, m \}$ and
suppose that all sets $Q(I)_{\alpha'}$ with 
$\alpha' \colon Z' \rightarrow \dom(I)$
and $Z'\in \binom{X}{m-(t-1)}$ have cardinality at most $(t-1)!^2\cdot k^{t-1}$. 

\smallskip
\noindent
{\em ``well-defined''.}
Let $\alpha$ be of the form $\alpha:Z\rightarrow\dom(I)$ with $Z\in \binom{X}{m-t}$ and assume that $|Q(I)_{\alpha}| > t!^2\cdot k^t$.
For arbitrary $\gamma\in Q(I)_{\alpha}$, we define the set 
$C_{\gamma}$ as 
\[C_{\gamma}:=\{\gamma' \in Q(I)_{\alpha} : \Delta(\gamma,\gamma') < t\},\]
i.e., $C_{\gamma}$ contains the solutions whose distance from $\gamma$ is less than $t$ or, 
equivalently, that agree with $\gamma$ on at least one variable from $X \setminus Z$.
Hence, we have
\[C_{\gamma} = \bigcup_{x\in X\setminus Z} Q(I)_{\alpha \cup \{x \mapsto \gamma(x)\}}\]
and thus, the size of $C_{\gamma}$ is at most $t\cdot(t-1)!^2\cdot k^{t-1}$
by the assumption of the lemma.

Now, iteratively select elements $\gamma_i$ for $i \in \{1, \dots,t\cdot k\}$ 
with $\gamma_i \in Q(I)_{\alpha} \setminus \bigcup_{j=1}^{i-1}C_{\gamma_j}$, i.e., 
arbitrarily choose $\gamma_1 \in Q(I)_{\alpha}$, 
then $\gamma_2 \in Q(I)_{\alpha} \setminus C_{\gamma_1}$, 
then $\gamma_3 \in Q(I)_{\alpha} \setminus (C_{\gamma_1} \cup C_{\gamma_2})$, etc.

We claim that such elements $\gamma_i$ for $i \in \{1, \dots,t\cdot k\}$ indeed exist, i.e., 
for every  $i \in \{1, \dots,t\cdot k\}$, $|Q(I)_{\alpha} \setminus \bigcup_{j=1}^{i-1}C_{\gamma_j}| > 0$. 
Indeed, by the assumption $|Q(I)_{\alpha}| \geq t!^2\cdot k^t$
and the above considerations on the size of $C_\gamma$ for arbitrary $\gamma$, 
we have: 
\[|Q(I)_{\alpha}\setminus \bigcup_{j=1}^{i-1}C_{\gamma_j}|\geq t!^2\cdot k^t - (i-1) \cdot t\cdot(t-1)!^2\cdot k^{t-1}  
> t!^2\cdot k^t - (t \cdot k) \cdot t\cdot(t-1)!^2\cdot k^{t-1}  = 0.
\]
Now set $\Gamma = \{\gamma_1,\dots,\gamma_{t\cdot k}\}\subseteq Q(I)_{\alpha}$.
By the construction, we have that $\gamma_i$ differs from $\gamma_j$ for $j < i$ on all variables $X\setminus Z$ as $\gamma_i\not \in C_{\gamma_j}$.
Hence, $\textbf{Red}_t$ is well-defined, i.e., the desired $t \cdot k$
solutions indeed exist. 

Moreover, the proof also demonstrates that the set $\Gamma$ can be constructed by
starting with one solution $\gamma_1$ and then iteratively adding arbitrary solutions
$\gamma_i$ that just need to differ from all solutions $\gamma_1, \dots, \gamma_{i-1}$
selected so far.

\medskip
\noindent
{\em ``safe''.}
Let $\Iold$ denote the database instance before applying 
$\textbf{Red}_t$ and let $\Inew$ denote its state after an application
of $\textbf{Red}_t$, i.e., $Q(\Inew) = (Q(\Iold) \setminus Q(\Iold)_\alpha) \cup \Gamma$. 
Now consider arbitrary pairwise distinct solutions 
$\gamma_1,\dots,\gamma_k \in Q(\Iold)$. 
We have to show that there exist pairwise distinct 
solutions $\gamma'_1, \dots, \gamma'_k$ in $Q(\Inew)$ with 
    $\delta(\gamma'_1, \dots, \gamma'_k) \geq \delta(\gamma_1, \dots, \gamma_k)$.

Assume that, for some $i \in \{1, \dots, k\}$, 
$\gamma_i$ gets removed by $\textbf{Red}_t$, 
i.e., $\gamma_i\in Q(\Iold)_\alpha \setminus \Gamma$.
We claim that there exists $\gamma'_i \in \Gamma \subseteq Q(\Inew)$
with
$\delta(\gamma_1, \dots, \gamma_{i-1}, \gamma'_i, \gamma_{i+1}, \dots,
\gamma_k) \geq \delta(\gamma_1, \dots, \gamma_k)$ and is different to $\gamma_1,\dots,\gamma_{i-1},\gamma_{i+1},\dots,\gamma_k$.

For arbitrary $j \neq i$, we define the set $\Gamma_j \subseteq \Gamma$
as 
$\Gamma_j =\{\gamma' \in \Gamma : \revc{\Delta(\gamma',\gamma_j) < \Delta(\gamma_i,\gamma_j)}\}$,
i.e., 
$\Gamma_j$ contains those
elements of $\Gamma$ whose distance from $\gamma_j$ 
is smaller than the distance between $\gamma_{i}$ and $\gamma_j$.
We will show below that
$|\Gamma_j| \leq t$ holds. 
In this case,  we have 
\[|\Gamma \setminus \bigcup_{i\neq j}\Gamma_j| \geq 
t \cdot k - t \cdot (k-1) = t \geq 1.
\]
That is, 
$\Gamma \setminus \bigcup_{i\neq j}\Gamma_j \neq \emptyset$. In other words, 
we can choose a solution $\gamma'_i$ from $\Gamma$ that differs from all $\gamma_j$ at least as much as $\gamma_i$ did.
Hence, such $\gamma'_i$ indeed has the property
$\delta(\gamma_1, \dots, \gamma_{i-1}, \gamma'_i, \gamma_{i+1},
\gamma_k) \geq \delta(\gamma_1, \dots, \gamma_k)$ and is different to $\gamma_1,\dots,\gamma_{i-1},\gamma_{i+1},\dots,\gamma_k$. 
By iterating this argument for every $i \in \{1, \dots, k\}$, we may conclude that 
there exist pairwise distinct solutions 
$\gamma'_1, \dots, \gamma'_k \in \Inew$ with 
$\delta(\gamma'_1, \dots, \gamma'_k) \geq \delta(\gamma_1, \dots, \gamma_k)$.

It only remains to show that $|\Gamma_j| \leq t$ indeed holds. 
As $\gamma_i$ and any element $\gamma' \in \Gamma\subseteq Q(\Iold)_{\alpha}$ agree on the variables $Z$, 
a lower diversity can only be achieved by $\gamma'$, if $\gamma_j$ and $\gamma'$
agree on some variable $x\in X\setminus Z$. 
We define 
\[
\Gamma_j^{(x)} = \{ \gamma' \in \Gamma : \gamma'(x) =  \gamma_j(x) \}.
\]
Hence, 
\[\Gamma_j \subseteq \bigcup_{x\in X\setminus Z} \Gamma_j^{(x)}.\]
Now, if some $\gamma'$ is in $\Gamma_j^{(x)}$, all other $\gamma''\in \Gamma, \gamma'\neq \gamma''$ are not in $\Gamma_j^{(x)}$ as $\gamma'$ and $\gamma''$ differ on $x\in X\setminus Z$ by construction of $\Gamma$.
Therefore, $|\Gamma_j^{(x)}|\leq 1$ and
\[|\Gamma_j|\leq \sum_{x\in X\setminus Z}|\Gamma_j^{(x)}|\leq |X\setminus Z|=t.\]
This completes the proof of the claim.
\end{proof}

\noindent
With the reduction rule $\textbf{Red}_t$ at our disposal, we can design an $\fpt$-algorithm 
(data complexity) for 
\DiverseACQmon and, more generally, for the \DiverseFOmon problem:

\begin{thm} \label{theorem:cqFPT}
    The problem \DiverseFOmon 
    is in $\fpt$ \revc{in} data complexity 
    when parameterized by the size $k$ of the diversity set.
    More specifically, an instance $\langle I, Q, k, d \rangle$ of \DiverseFOmon with $m$-ary FO-query~$Q$ can be reduced in 
    polynomial time (data complexity) to an equivalent instance $\langle I', Q', k, d \rangle$
    of \DiverseFOmon
    of size $\O(m!^2\cdot k^m)$.
\end{thm}
\begin{proof}
Recall that we may assume that query $Q$ is of the form 
$Q(x_1,\dots,x_m):=  R(x_1,\dots,x_m)$
and the database $I$ consists of a single relation $R^I$.
We apply $\textbf{Red}_1$ through $\textbf{Red}_m$ to $I$ in this order exhaustively.
\revc{Initially, we have to check the preconditions of \autoref{lem:claimA} for $t=1$ for us to safely apply $\textbf{Red}_1$.
Thus, let us consider $Z\in \binom{X}{m}$.
We have $Z=X$ and hence, for every $\alpha:Z\rightarrow\dom (I)\in Q(I)$, 
we have $Q(I)_{\alpha} = \{\alpha\}$. 
In particular, 
$|Q(I)_{\alpha}| = 1 \leq 0 \cdot k + k^0$. 
Hence, 
the preconditions of \autoref{lem:claimA} are fulfilled and exhaustive application of 
$\textbf{Red}_1$ does not alter the status of the \DiverseFOmon problem. 
}
After exhaustive application of $\textbf{Red}_1$, if now $\textbf{Red}_2$ is applicable, then 
the preconditions of \autoref{lem:claimA}  are fulfilled and exhaustive application of 
$\textbf{Red}_2$ does not alter the status of the \DiverseFOmon problem, etc.

Finally, after exhaustive application of 
$\textbf{Red}_m$, let $\Istar$ denote the resulting database instance.  
Note that, for $t = m$, we have
$\binom{X}{0}:= \{Z \subseteq X : |Z| = 0\} = \{ \emptyset\} $. 
and  $|Q(\Istar)_{\alpha}|\leq m!^2\cdot k^m$ for any $\alpha: \emptyset \rightarrow \dom(\Istar)$. 
In particular, this means that such an $\alpha$ does not bind any variables in $X$. 
Hence, $Q(\Istar)_\alpha = Q(\Istar)$ and, therefore, 
$|Q(\Istar)|\leq m!^2\cdot k^m$. By the form of $Q$ (with a single atom) and $\Istar$ (with a single relation), 
this means $\Istar$ is of size $\O(m!^2\cdot k^m)$.

It remains to show that the exhaustive application of $\textbf{Red}_1$ through $\textbf{Red}_m$
is feasible in polynomial time data complexity.
In total, we have to consider at most $2^m$ sets $Z \subseteq X$ of variables
with $|Z| = m-t$ for $t \in \{1, \dots, m\}$ and check if $\textbf{Red}_t$ is applicable. 

For each $Z$, if the reduction rule  is applicable, 
the following computation is carried out. 
Let $Z= \{z_1, \dots, z_{m-t}\}$ and $X \setminus  Z= \{z_{m-t+1}, \dots, z_{m}\}$. 
Moreover, let $S \subseteq R^I$ denote the subset of answer tuples that are still left after 
previous applications of the reduction rule. 
Then we order $S$ lexicographically for this variable order. 
That is, tuples with the same value combination on 
$Z$ occur in contiguous positions. In a single pass of the ordered instance $S$ we 
inspect, for each  value combination $\alpha$ on $Z$, the set $S_\alpha \subseteq S$ of tuples with 
precisely this value combination $\alpha$ on $Z$. 
If $|S_\alpha| <  t!^2 \cdot k^t$, then we do nothing. 
Otherwise, we select $t \cdot k$ tuples from $S_\alpha$. By
the last property of \autoref{lem:claimA}, we can apply the following
steps: 
choose the first tuple $\gamma_1 \in S_\alpha$;  then, for every 
$i \in \{2, \dots, t \cdot k\}$, 
further scan $S_\alpha$ until a tuple 
$\gamma_i \in S_\alpha$ is found that differs from all tuples 
$\gamma_1, \dots, \gamma_{i-1}$ on all variables $X \setminus Z$.
Since \autoref{lem:claimA} guarantees that we can just pick suitable solutions
in an arbitrary order, this approach is guaranteed to produce the required result.
Let $\Gamma = \{\gamma_1, \dots, \gamma_{t \cdot k}\}$. We may then delete all tuples in 
$S_\alpha \setminus \Gamma$ from $S$.

The total effort for the exhaustive application of the reduction rule $\textbf{Red}_t$ for $t \in \{1, \dots, m\}$ is obtained by the following considerations: 

\begin{itemize}
\item   Evaluating the original, general FO-formula over the original database instance is 
feasible in polynomial time data complexity. Also, the size of the resulting answer relation $R^ I$ is of course bounded by this polynomial. Let us denote it by $p$.

    \item There is an ``outer loop'' over subsets 
    $Z \subseteq X$. There are $2^m$ subsets, where $m$ depends 
    only on the query, which is considered as constant in data complexity. 
    \item Inside this loop, we first sort the set of remaining 
    answer tuples $S \subseteq R^ I$. The effort for this step 
    is bounded by $\O(p \cdot \log(p)\cdot m)$.
    \item One pass of (the ordered set) $S$ has cost
    $\leq p$.
    \item For each $S_\alpha$, we check if 
    $|S_\alpha| \geq  t!^2 \cdot k^t$. If this is the case, 
    we select $t \cdot k$ tuples from  $S_\alpha$ in a single pass of $S_\alpha$. 
    This step is feasible in time $\O(p \cdot t \cdot k \cdot m)$ -- including also the cost for checking if the currently scanned tuple in $S_\alpha$ differs on all variables in $X \setminus Z$ 
    from the already selected tuples $\gamma_i$. 
    \item The deletion of the tuples in $S_\alpha \setminus \Gamma$ from $S$ can be done by first of all marking them as deleted when 
    constructing the set $\Gamma$. When all $\alpha$'s have been processed, we can actually delete the marked tuples from $S$ by yet another pass of $S$, which clearly fits into $\O(p)$ time. \qedhere
\end{itemize}
\end{proof}

\noindent
We now study the data complexity of the \DiverseACQ problem in the non-parameterized case, i.e., the size 
$k$ of the diversity set is part of the input and no longer considered as the parameter. It will turn out that this problem is $\np$-hard
for any ws-monotone diversity measure.
Our $\np$-hardness proof will be by reduction from the \IndependentSet problem, where we restrict the instances to graphs of degree at most $3$. 
It was shown in~\cite{DBLP:conf/ciac/AlimontiK97} 
that this restricted problem remains $\np$-complete.

\begin{thm} \label{theorem:acqData}
 The problem \DiverseACQstrmon is $\np$-hard \revc{in} data complexity.
 It is  $\np$-complete if
 the size $k$ of the diversity set is given in unary.
\end{thm}
\begin{proof}
The $\np$-membership is immediate: compute $Q(I)$ (which is feasible in polynomial time 
when considering the query as fixed), then guess a subset $S \subseteq Q(I)$ of size $k$ and check in polynomial time that $S$ has the desired diversity. 

We prove hardness by reduction from a restricted version of the \IndependentSet problem, where 
we assume all instances of graphs to be of degree at most $3$. Before we present the reduction, 
let us briefly look at this restriction. 

It is easy to see that \IndependentSet remains $\np$-hard
if we restrict the degree of vertices to 4. This result is obtained 
by combining two classical results~\cite{Papadimitriou1994}: 
first, \textsc{3-SAT} remains $\np$-hard even if every variable 
in the propositional formulas occurs at most 3 times and each literal 
occurs at most 2 times. And second, we apply the reduction from 
\textsc{3-SAT}  to 
\IndependentSet where each clause is represented by a triangle and any
two dual literals are connected by an edge. Hence, in the resulting graph, 
each vertex is adjacent two at most 4  vertices (the other 2 vertices
in the triangle plus at most 2 vertices corresponding to dual literals). 
This result was strengthened in 
\cite{DBLP:conf/ciac/AlimontiK97} 
where it was shown that we may even further restrict the degree of vertices to 
3. The idea of Alimonnti and Kann 
\cite{DBLP:conf/ciac/AlimontiK97} is to apply the following transformation 
for each vertex of degree greater than 3: 
suppose that $v$ has degree greater than 3; then replace $v$ 
by a path $v_1,v_2,v_3$,
where 2 edges containing $v$ are connected to $v_1$ and the remaining edges of $v$ 
are connected to $v_3$.
Thus, $v_1$ and $v_2$ have degree less than or equal to 3 while the degree of $v_3$ is strictly less than the degree of $v$.
Furthermore, the original graph has an independent set of size $k$ if and only if the new one has an independent set of size $k+1$ as picking $v_1$ and $v_3$ corresponds to picking $v$. Exhaustive application of this transformation yields an instance
of \IndependentSet where every vertex in the graph has degree $\leq 3$.

For the $\np$-hardness, we define the 
query $Q$ independently of the instance of the \IndependentSet problem 
as 
$Q(x_1,x_2,x_3,x_4,x_5) :=  R(x_1,x_2,x_3,x_4,x_5)$, i.e., the only relation symbol $R$ has arity 5.
Now let $(G,s)$ be an instance of \IndependentSet where each vertex of $G$ has degree at most 3.

Let $V(G)=\{v_1,\dots,v_n\}$ and $E(G)=\{e_1,\dots,e_m\}$.
Then the database $I$ consists of a single relation $R^I$ with $n$ tuples (= number of vertices in $G$) over the domain $\dom(I)=\{\textbf{free}_1,\dots,\textbf{free}_n,\textbf{taken}_1,\dots,\textbf{taken}_m\}$.
The $i$-th tuple in $R^I$ will be denoted $(e_{i,1},\dots,e_{i,5})$. 
For each $v_i\in V(G)$, the values $e_{i,1},\dots,e_{i,5}\in\dom (I)$ are defined by an iterative process:

\begin{enumerate}
    \item The iterative process starts by initializing all $e_{i,1},\dots,e_{i,5}$ to $\textbf{free}_i$ for each $v_i\in V(G)$.
    \item We then iterate through all edges $e_j\in E(G)$ and do the following:  
    Let $v_{i}$ and $v_{i'}$ be the two incident vertices to $e_j$
    and let $t\in\{1,\dots,5\}$ be an index such that $e_{i,t}$ and $e_{i',t}$ both still have the values $\textbf{free}_{i}$ and $\textbf{free}_{i'}$, respectively.
    Then set both $e_{i,t}$ and $e_{i',t}$ to $\textbf{taken}_j$.
\end{enumerate}

\noindent
In the second step above when processing an edge $e_j$, such an index $t$ must always exist. 
This is due to the fact that, at the moment of considering $e_j$, the vertex $v_{i}$ has been considered at most twice (the degree of $v_{i}$ is at most 3) and thus, for at least three different values of $t\in \{1,\dots,5\}$, the value $e_{i,t}$ is still set to $\textbf{free}_{i}$. 
Analogous considerations apply to vertex $v_{i'}$ and thus, for at least 3 values of $t\in \{1,\dots,5\}$, we have $e_{i',t} = \textbf{free}_{i'}$. 
Hence, by the pigeonhole principle, there exists $t\in \{1,\dots,5\}$ with $e_{i,t} = \textbf{free}_{i}$ and $e_{i',t} = \textbf{free}_{i'}$.

After the iterative process, the database $I$ is defined by $R^I=\{(e_{i,1},e_{i,2},e_{i,3},e_{i,4},e_{i,5}) : i=1,\dots,n\}$.
Moreover, the size of the desired diversity set is set to $k=s$
and the target diversity is set to 
$d = f(5, \dots, 5)$, where $f$ is the aggregator of \revc{$\deltastrmon$ aggregating the value $5$ exactly $\binom{k}{2}$ times}.
The resulting problem instance is of the form 
$\langle I,Q,k,d\rangle$.

The reduction is clearly feasible in polynomial time.
Its correctness, i.e., the graph $G = (V(G), E(G))$ having an independent set of 
size $s$ if and only if there exists $S \subseteq Q(I)$ with $|S| = k$ and diversity 
$\geq d$ hinges on the observation 
that the desired diversity can only be reached by $k$ answer tuples that pairwisely differ
in all 5 positions due to ws-monotonicity.
Furthermore, the answers $Q(I)$ are trivially $\{\gamma_1,\dots,\gamma_n\}$ with $\gamma_i(x_t)=e_{i,t}$
for each $t\in\{1,\dots,5\}$. We thus have to show that these differ on all values if and 
only if $G$ has an independent set of size $s = k$.

First, suppose that $G$ has such an independent set, say
$\{v_{i_1}, \dots, v_{i_k}\}$. We claim that then 
$\{\gamma_{i_1}, \dots, \gamma_{i_k}\}$ is a subset of $Q(I)$ with the 
desired diversity, i.e., any two answers $\gamma_{i_r}$ and $\gamma_{i_s}$
differ on all 5 variables. Suppose to the contrary that 
$\gamma_{i_r}(t) = \gamma_{i_s}(t)$ holds for some $t \in \{1, \dots, 5\}$. 
By our construction of $R^I$, this can only happen if 
$\gamma_{i_r}(t) \neq \textbf{free}_{i_r}$ and
$\gamma_{i_s}(t) \neq \textbf{free}_{i_s}$. Hence, 
$\gamma_{i_r}(t) = \gamma_{i_s}(t) = \textbf{taken}_j$  for some $j \in \{1, \dots, m\}$ holds.
Again by our construction of $R^I$, this means that both 
$v_{i_r}$ and $v_{i_s}$ are incident to the edge $e_j$. 
This contradicts the assumption that both $v_{i_r}$ and $v_{i_s}$ are
contained in an independent set.

Conversely, suppose that there exists a subset $S \subseteq Q(I)$ of size 
$k$ with the desired target diversity. 
Let $S = \{\gamma_{i_1}, \dots, \gamma_{i_k}\}$. 
We claim that then $\{v_{i_1}, \dots, v_{i_k}\}$ is an independent set of $G$. 
Suppose to the contrary that it is not, i.e., 
two vertices  $v_{i_r}$ and $v_{i_s}$ are incident to the same edge $e_j$. 
Then, by our construction of $I$, 
there exists $t \in \{1, \dots, 5\}$ with 
$\gamma_{i_r}(t) = \gamma_{i_s}(t) = \textbf{taken}_j$. 
This means that $\Delta(\gamma_{i_r},\gamma_{i_s})<5$ and hence, the target diversity
$f(5, \dots, 5)$ cannot be reached by $S$ due to ws-monotonicity, which is a contradiction.
\end{proof}

\section{Diversity of Unions of Conjunctive Queries}
\label{sec:UCQs}

We now turn our attention to UCQs.
Of course, 
all hardness results proved for CQs and ACQs in 
Section~\ref{sec:CQs} carry over to UCQs and UACQs, respectively. 
Moreover, the $\fpt$-membership result from 
\autoref{theorem:cqFPT} for general FO-formulas of course also 
includes UCQs. 
It remains to study the query complexity and combined complexity 
of UACQs. It turns out that 
the union makes the problem significantly harder than for ACQs \revc{and we are not able to establish $\xp$-membership}. 
Instead, we show next that \DiverseUACQstrmon is $\np$-hard
even in a very restricted setting, namely \revc{where we are looking for a pair of diverse answers to a union of two ACQs over a fixed database.
Put differently, \DiverseUACQstrmon is $\np$-hard in query complexity even when fixing the parameter to $k=2$, making the existence of an $\xp$-algorithm unlikely.}

The proof will be by reduction from a variant of the 
\textsc{List Coloring} problem,  
which we introduce next:
A \emph{list assignment} $C$ assigns each vertex $v$ of a graph $G$ a list of colors $C(v)\subseteq\{1,\dots,l\},l\in \mathbb{N}$.
Then a \emph{coloring} is a function $c:V(G)\rightarrow \{1,\dots,l\}$ and it is called $C-$\emph{admissible} if each vertex $v\in V(G)$ is colored in a color of its list, i.e., $c(v)\in C(v)$, and adjacent vertices $u, v\in E(G)$ are colored with different colors, i.e., $c(u)\neq c(v)$. Formally, the problem is defined as follows:

\begin{problem}{\textsc{List Coloring}}
    Input: A graph $G$, an integer $l\in\mathbb{N}$, and a list assignment $C:V(G)\rightarrow 2^{\{1,\dots,l\}}$.

    Question: Does there exist a $C$-admissible coloring $c:V(G)\rightarrow \{1,\dots,l\}$?
\end{problem}

\noindent
Clearly, \textsc{List Coloring} is a generalization of \textsc{3-Colorability} and, hence, 
$\np$-complete.
It was shown in \cite{DBLP:journals/dam/ChlebikC06}, that the \textsc{List Coloring} problem remains $\np$-hard even 
when assuming that each vertex of $G$ has degree 3, 
$G$ is bipartite, and $l=3$.
This restriction will be used in the proof of the 
following theorem.
\revc{Note that these restrictions also imply that both parts of the bipartition are of the same size.}

\begin{thm} \label{theorem:uacq}
    The problem \DiverseUACQstrmon is $\np$-hard \revc{in}
    query complexity 
    (and hence, also \revc{in} combined complexity). 
    It remains $\np$-hard even if the desired size of the diversity set is bounded by 2 and the UACQs are restricted to containing at most 
    two CQs and no existential variables.
    The problem is $\np$-complete if 
    the size $k$ of the diversity set is given in unary.
\end{thm}

\begin{proof}
The $\np$-membership in case of $k$ given in unary is immediate: 
guess $k$ assignments to the free variables of query $Q$, check in polynomial time that they are solutions, 
and verify in polynomial time that their diversity is above the 
desired threshold.

For our problem reduction, we consider a fixed database $I$ over a fixed schema, which 
consists of the 4 domain elements $\dom(I)=\{0,1,2,3\}$ and 
9 relation symbols 
\[R_{\{1\}}, R_{\{2\}}, R_{\{3\}},
 R_{\{1,2\}},  R_{\{1,3\}},  R_{\{2,3\}},  
 R_{\{1,2,3\}}, S, S'.
\]
The relations of the database are defined as follows:
\begin{align*}
    R^I_{\{1\}} &= \{(1,1,1)\}, \quad\quad\quad
\quad\quad\quad\quad\quad \ \
      R^I_{\{1,2\}} = \{(1,1,1), (2,2,2)\},
\\
    R^I_{\{2\}} &= \{(2,2,2)\}, \quad\quad\quad
\quad\quad\quad\quad\quad \ \
      R^I_{\{1,3\}} = \{(1,1,1), (3,3,3)\},
\\      
    R^I_{\{3\}} &= \{(3,3,3)\}, \quad\quad\quad
    \quad\quad\quad\quad\quad \ \ 
      R^I_{\{2,3\}} = \{(2,2,2), (3,3,3)\},
\\      
    R^I_{\{1,2,3\}} &= \{(1,1,1), (2,2,2),(3,3,3)\},
    \quad\quad\quad
    S^I =  \{(0)\},
        \quad\quad\quad \quad
    S'^I =  \{(1)\}.
\end{align*}

\noindent
Now let $\langle{}G,l,C\rangle$ be an arbitrary instance of 
\textsc{List Coloring}, where each vertex of $G$ has degree 3, $G$ is bipartite, and $l=3$.
That is, $G$ is of the form $G=(V\cup V',E)$ for vertex sets 
$V,V'$ and edge set $E$ with 
$V = \{v_1,\dots,v_n\}$, 
$V'=\{v'_1,\dots,v'_n\}$, and 
$E=\{e_1,\dots,e_{3n}\}$.
Note that $|V|=|V'|$ and $|E| = 3 \cdot |V|$ 
as each vertex in $G$ has degree 3 and $G$ is bipartite.

From this, we construct a UACQ $Q$ as follows: 
we use the $3n+1$ variables $x_1,\dots,x_{3n},y$ in our query.
For each $i \in \{1, \dots, n\}$,  
we write 
$e_{j_{i,1}},e_{j_{i,2}},e_{j_{i,3}}$ to denote the 
three edges incident to the vertex $v_i$.
Analogously, 
we write 
$e_{j'_{i,1}},e_{j'_{i,2}},e_{j'_{i,3}}$ 
to denote the 
three edges incident to the vertex $v'_i$.

The UACQ $Q$ is then defined as 
$Q (x_1,\dots ,x_{3n},y) :=  \phi \lor \psi$
with
\begin{align*}
    \phi &= \bigwedge_{i=1}^n R_{C(v_i)} (x_{j_{i,1}},x_{j_{i,2}},x_{j_{i,3}})  \land S (y),\\
    \psi &= \bigwedge_{i=1}^n R_{C(v'_i)} (x_{j'_{i,1}},x_{j'_{i,2}},x_{j'_{i,3}})
    \land S' (y).
\end{align*}
Moreover, we set the target diversity to $d=f(3n+1)$, where $f$ is the aggregator of $\delta$, and we are looking for $k=2$ solutions to reach this diversity.
Observe that each variable appears exactly once in $\phi$ and once in $\psi$, which makes both formulas trivially acyclic.
Furthermore, $Q$ contains no existential variables.

The intuition of the big conjunction in $\phi$ (resp.\ $\psi$)
is to ``encode'' for each vertex $v_i$ (resp.\ $v'_i$)
the 3 edges incident to this vertex in the form of the 3 $x$-variables with the corresponding indices. 
The relation symbol chosen for each vertex $v_i$ or $v'_i$ depends on the color list for this vertex. For instance, 
if $C(v_1) = \{2,3\}$ and if $v_1$ is incident to the edges 
$e_4, e_6, e_7$, 
then the first conjunct in the definition 
of $\phi$ is of the form 
$R_{\{2,3\}} (x_4,x_6,x_7)$.
Note that the order of the variables in this atom is irrelevant since the $R$-relations contain only tuples with identical values in all 3 positions. 
Intuitively, this ensures that a vertex (in this case $v_1$) gets the same color (in this case color 2 or 3) in all its incident edges (in this case $e_4, e_6, e_7$).  

It remains to prove the correctness of this reduction.
For this, 
observe that diversity $d = f(3n+1)$ can only be achieved by two answers $\gamma,\gamma'$ that differ on all variables due to ws-monotonicity.
Due to the $y$ variable with possible values 0 and 1, 
one answer has to satisfy $\phi$ while the other answer satisfies $\psi$.
W.l.o.g., let $\gamma$ satisfy $\phi$ and let 
$\gamma'$ satisfy $\psi$.
The intuition behind the reduction is that $\gamma$ tells us how to color the vertices in $V$ 
while $\gamma'$ tells us how to color the vertices in $V'$.

We have to show that $(G, l,C)$ is a 
positive instance of \textsc{List Coloring} if and only if 
$\langle{}Q,I, 2, f(3n+1)\rangle$ is a positive instance of \DiverseUACQstrmon.

\smallskip
For the ``only if''-direction, 
suppose  that $(G, l,C)$ is a positive instance of \textsc{List Coloring}, i.e., 
graph $G$ has a $C$-admissible coloring $c:V \cup V'\rightarrow \{1,2,3\}$. 
From this, we construct the assignments $\gamma$ and $\gamma'$ to the 
$3n + 1$ variables in $Q$ as follows: 

\smallskip
\noindent
$\gamma(y) = 0$ and
$\gamma(x_{j_{i,1}})=\gamma(x_{j_{i,2}})=\gamma(x_{j_{i,3}})=c(v_i)$
for every $i \in \{1, \dots, n\}$  
and, analogously, \\
$\gamma'(y) = 1$ and
$\gamma'(x_{j'_{i,1}})=\gamma'(x_{j'_{i,2}})=\gamma(x_{j'_{i,3}})=c(v'_{i})$
for every $i \in \{1, \dots, n\}$.

\smallskip
\noindent
We first have to verify that $\gamma$ is a solution of $\phi$ and
$\gamma'$ is a solution of $\psi$. We only do this for $\gamma$. The argumentation for 
$\gamma'$ is analogous. $S(\gamma(y)) = S(0)$ is clearly contained in database $I$. 
Now consider an arbitrary index $i \in \{1, \dots, n\}$. 
The atom $R_{C(v_i)} (x_{j_{i,1}},x_{j_{i,2}},x_{j_{i,3}})$ is sent to 
$R_{C(v_i)} (c(v_i),c(v_i),c(v_i))$  by $\gamma$.
By the above construction of database $I$, the tuple 
$(c(v_i),c(v_i),c(v_i))$ is indeed contained in relation $R_{C(v_i)}^I$.

It remains to show that the two assignments $\gamma$ and $\gamma'$ differ on every variable. 
Let $x_{j_{r,t}}$ and $x_{j'_{s,u}}$ with $r,s \in \{1, \dots, n\}$ and $t,u \in\{1,2,3\} $
denote the same variable. By our construction of the $R$-atoms in $\phi$ and $\psi$, 
this means that 
$e_{j_{r,t}}$ and $e_{j'_{s,u}}$ denote the same edge in $G$ and
$v_r$ and $v'_s$ are the two endpoints of this edge. 
Since $c$ is a $C$-admissible coloring, we have $c(v_r) \neq c(v'_s)$. 
Moreover,  by our definition of $\gamma$ and $\gamma'$, we have 
$\gamma(x_{j_{r,t}}) = c(v_r)$ and $\gamma'(x_{j'_{s,u}}) = c(v'_s)$. 
Hence, $\gamma$ and $\gamma'$ indeed differ on an arbitrarily chosen variable and, thus,
on every variable. 

\smallskip
For the ``if''-direction, 
suppose  that $\langle{}Q,I, 2, f(3n+1)\rangle$ is a positive instance of \DiverseUACQ,
i.e., there exist two solutions $\gamma$ and $\gamma'$ with diversity $f(3n + 1)$. 
This means that $\gamma$ and $\gamma'$ differ on every variable, in particular on $y$. Hence, 
one of the solutions is an answer of $\phi$ and one of $\psi$. W.l.o.g., let $\gamma$ 
be an answer of $\phi$ and let $\gamma'$ be an answer of $\psi$.
From this, we construct the following coloring $c:V \cup V'\rightarrow \{1,2,3\}$:

\smallskip
\noindent
$c(v_i) = \gamma (x_{j_{i,1}})$ and 
$c(v'_i) = \gamma'(x_{j'_{i,1}})$
for every $i \in \{1, \dots, n\}$.

\smallskip
\noindent
We have to show that $c$ is $C$-admissible. 
Consider an arbitrary edge $e$ with endpoints $v_r$ and $v_s$
for $r,s \in \{1, \dots, n\}$.
By our construction of $Q$, there exist indices $t,u \in \{1,2,3\}$, such that 
$x_{j_{r,t}}$ and $x_{j'_{s,u}}$ 
denote the same variable. 
Since $\gamma$ and $\gamma'$ have diversity $f(3n+1)$, the assignments 
$\gamma$ and $\gamma'$ differ on every variable. In particular, 
we have
$\gamma(x_{j_{r,t}})  \neq  \gamma'(x_{j'_{s,u}})$. 
Moreover, by our definition of coloring $c$ and the database $I$, we have 
$c(v_r) = \gamma(x_{j_{r,1}})= \gamma(x_{j_{r,t}})$ and $c(v'_s) = \gamma'(x_{j'_{s,1}}) = \gamma'(x_{j'_{s,u}})$. 
Hence, $c$ assigns different colors to the two arbitrarily chosen, adjacent vertices 
$v_r$ and $v'_s$ and, therefore, to any adjacent vertices of $G$. That is, 
$c$ is $C$-admissible.
\end{proof}

\section{Diversity of Conjunctive Queries with Negation}
\label{sec:CQneg}

Lastly, we consider \cqsn.
As was recalled in Section~\ref{sec:introduction}, the restriction to acyclicity is not sufficient to ensure tractable answering of 
\cqsn~\revc{\cite{DBLP:journals/tcs/Lanzinger23}}. 
In the following, we thus  
restrict ourselves to queries of bounded treewidth
when analyzing the \DiverseCQn problem.

The data complexity case has already been settled for arbitrary FO-formulas in \autoref{theorem:cqFPT}. Hence, of course, also  \DiverseCQn is in $\fpt$ data complexity and $\np$-hard in the non-parameterized case. Moreover, we observe that the query used in the proof of
\autoref{theorem:acqCCHardness} has a treewidth of one. Hence, it is clear that also \DiverseCQnstrmon is $\wone$-hard combined complexity for queries with bounded treewidth. It remains to study the combined complexity upper bound, for which we 
describe an  $\xp$-algorithm next.

Our algorithm is based on so-called {\em nice} tree decompositions -- a normal form introduced in \cite{DBLP:books/sp/Kloks94}. 
A nice tree decomposition only allows leaf nodes plus three types of inner nodes: introduce nodes, forget nodes, and join nodes.
An \textit{introduce node} $t$ has a single child $t'$ with $\chi(t) = \chi(t') \cup \{z\}$ for a single variable $z$.
Similarly, a \textit{forget node} $t$ has a single child $t'$ with $\chi(t') = \chi(t) \cup \{z\}$ for a single variable $z$.
Finally, a \textit{join node} $t$ has two child nodes $t_1,t_2$ with $\chi(t) = \chi(t_1) = \chi(t_2)$. It was shown in 
\cite{DBLP:books/sp/Kloks94} that every tree decomposition can be transformed in linear time into a nice tree decomposition
without increasing the width. 

The intuition of the present algorithm is very similar to the intuition of \autoref{algo:one} presented in Section~\ref{sec:CQ-algorithm}.
That is, both algorithms maintain information on tuples of $k$ partial solutions in a set $D_t$.
Concretely, these tuples are again of the form $(\alpha_1,\dots,\alpha_k,(d_{i,j})_{1\leq i < j \leq k})$.
This time, however, partial solutions $\alpha_i$ are not assignments that satisfy concrete atoms but arbitrary assignments defined on $\chi(t)$.
Nevertheless, a tuple gets added to $D_t$ if and only if it is possible to extend the partial solutions to mappings $\gamma_1,\dots,\gamma_k$ that (a) satisfy the query associated to the subtree rooted in $t$ and (b) for $1\leq i<j\leq k$ the distance between $\gamma_i$ and~$\gamma_j$ is exactly $d_{i,j}$.

Formally, for a \cqn $Q(X) := \exists Y \bigwedge_{i=1}^n L_i(X,Y)$ and 
nice tree decomposition $\langle T, \chi, r \rangle$ of~$Q$ we define for $t\in V(T)$ the subquery
\[Q_{t}=\bigwedge_{\substack{i=1,\dots,n\\ \var(L_i)\subseteq \chi(t)}} L_i,\]
i.e., $Q_t$ contains those literals of~$Q$ whose variables are covered
by $\chi(t)$.

\begin{algo} \label{algo:two}
Given $Q(X)$, $I$, $k$, $d$, a nice tree decomposition $\langle{}T, \chi, r\rangle$
of minimum width, and a diversity measure $\delta$ defined via some
aggregate function $f$, the algorithm proceeds in two main steps: 
First, sets $D_t$ are computed bottom-up for each $t\in V(T)$, and then, it is determined from $D_r$
whether the diversity threshold $d$ can be met.
For the bottom-up step, the type of $t$ determines how $D_t$ is computed:

\begin{itemize}
    \item \textbf{Leaf Node:} For a leaf node $t\in V(T)$ we create $D_t$ as
    \begin{align*}
    D_{t}=\{(\alpha_1,\dots,\alpha_k,(d_{i,j})_{1\leq i < j \leq k}) :{}&\alpha_1,\dots,\alpha_k\colon \chi(t)\rightarrow \dom(I), \nonumber \\
    &\alpha_1,\dots,\alpha_k \text{ satisfy } Q_{t},\nonumber \\  
    &  d_{i,j} = \Delta_X(\alpha_i,\alpha_j),1\leq i<j\leq k\}.
    \end{align*}
    Hence, we exhaustively go through all possible variable assignments $\alpha_1,\dots,\alpha_k\colon \chi(t)\rightarrow \dom(I)$, keep those which satisfy the query $Q_t$, and record their pairwise diversities.
    \item \textbf{Introduce Node:} For an introduce node $t\in V(T)$ with child 
    $c\in V(T)$ which introduces the variable $z\in \chi(t)\setminus\chi(c)$, we create $D_t$ as
    \begin{align*}
    D_{t}=\{(\alpha_1 \cup \beta_1,\dots,\alpha_k \cup \beta_k,(d'_{i,j})_{1\leq i < j \leq k}) :{}&(\alpha_1,\dots,\alpha_k,(d_{i,j})_{1\leq i < j \leq k}) \in D_c, \nonumber \\
    &\beta_1,\dots,\beta_k\colon \{z\}\rightarrow \dom(I), \nonumber\\
    & \alpha_1\cup \beta_1,\dots, \alpha_k\cup \beta_k \text{ satisfy } Q_{t}, \nonumber \\
    & d'_{i,j} = d_{i,j} +  \Delta_X(\beta_i,\beta_j), 1\leq i < j \leq k \}.
    \end{align*}
    Thus, we extend the domain of the local variable assignments in $D_c$ by $z$.
    We do this by exhaustively going through all $e\in D_{c}$ in combination with all $\beta_1,\dots,\beta_k\colon \{z\}\rightarrow \dom(I)$, check if the extensions $\alpha_1\cup \beta_1,\dots, \alpha_k\cup \beta_k$ satisfy all literals for which all variables are covered, and, if this is the case, add the diversity achieved on the $z$-variable.
    \item \textbf{Forget Node:} For a forget node $t\in V(T)$ with child $c\in V(T)$ we create $D_t$ as
    \begin{align*}
        D_{t}=\{(\alpha_1|_{\chi(t)},\dots,\alpha_k|_{\chi(t)},(d_{i,j})_{1\leq i < j \leq k}) :{}&(\alpha_1,\dots,\alpha_k,(d_{i,j})_{1\leq i < j \leq k})\in D_c\}.
    \end{align*}
    
    \item \textbf{Join Node:} For a join node $t\in V(T)$ with children $c_1,c_2\in V(T)$ we create $D_t$ as
    \begin{align*}
        D_{t}=\{(\alpha_1,\dots,\alpha_k,(d_{i,j})_{1\leq i < j \leq k}) :{}&(\alpha_1,\dots,\alpha_k,(d'_{i,j})_{1\leq i < j \leq k}) \in D_{c_1}, \nonumber \\
        &(\alpha_1,\dots,\alpha_k,(d''_{i,j})_{1\leq i < j \leq k}) \in D_{c_2}, \nonumber \\
        & d_{i,j} = d'_{i,j} + d''_{i,j} -  \Delta_X(\alpha_i,\alpha_j), 1\leq i < j \leq k \}.
    \end{align*}
    In this step, we match rows of $D_{c_1}$ with rows of $D_{c_2}$ that agree on the local variable assignments and simply combine the diversities achieved in the two child nodes while subtracting the diversity counted twice. 
\end{itemize}

For the second step, the algorithm goes through all $(\alpha_1,\dots,\alpha_k,(d_{i,j})_{1\leq i < j \leq k})\in D_r$ and removes those tuples where $d_{i,j}=0$ for at least one $1\leq i < j \leq k$ or $f((d_{i,j})_{1\leq i < j \leq k}) < d$.
Then, the algorithm returns ``yes'' if the resulting set is non-empty and otherwise ``no''.
\end{algo}

Clearly, the algorithm is well-defined and terminates. 
The next theorem states that the algorithm decides \DiverseCQn, and discusses its running time.

\begin{thm}\label{theorem:cqnXP}
 For a class of \cqsn of bounded treewidth, the problem \DiverseCQn is in $\xp$ \revc{in combined complexity}
 when parameterized by the size $k$ of the diversity set.
 More specifically, let $Q(X)$ be from a class of \cqsn which have treewidth $\leq \omega$.
 Then, for a database instance $I$ and integers~$k, d$, \autoref{algo:one} solves \DiverseCQn in time 
 $\O(\dom(I)^{2\cdot k\cdot (\omega + 1)}\cdot (|X|+1)^{k(k-1)} \cdot \pol(|Q|,k))$, 
 where $\pol(|Q|,k)$ is a polynomial in $|Q|$ and $k$.
\end{thm}
To prove this statement, we show by a sequence of lemmas that $D_t$ truly captures the intended meaning.
As before, let $Q(X)$ be a \cqn, $\langle T, \chi, r\rangle$ a nice tree decomposition of~$Q$, 
$k$ the number of elements in the diversity set, and $d$ the required diversity.
Furthermore, we extend the definition of $\chi$ and $Q_t$ to subtrees $T_t$ of $T$ rooted in $t$.
To that end, let $\chi(T_t)= \bigcup_{t'\in V(T_t)}\chi(t')$ and
\[Q_{T_t} = \bigwedge_{\substack{i=1,\dots,n \\
\var(L_i)\subseteq \chi(T_t)}}L_i.\]
With this, for a tuple
\[e=(\alpha_1,\dots,\alpha_k,(d_{i,j})_{1\leq i < j \leq k})\in (\chi(t)\rightarrow \dom(I))^k\times \{0,\dots,|X|\}^{\frac{k(k-1)}{2}}\]
we define a set of witnesses
\begin{align*}
    \wit{t}(e)=\{(\gamma_1,\dots,\gamma_k) :{}&\gamma_1,\dots,\gamma_k\in Q_{T_t}(I),\\
    & \gamma_1\cong \alpha_1,\dots,\gamma_k\cong\alpha_k,\\
    & d_{i,j}= \Delta_X(\gamma_i,\gamma_j),1\leq i < j \leq k\}.
\end{align*}
The existence of such extensions $(\gamma_1,\dots,\gamma_k)\in \wit{t}(e)$ are precisely guaranteed by $e\in D_t$ as they satisfy the query corresponding to the subtree rooted in $t$ and $(d_{i,j})_{1\leq i < j \leq k}$ are their pairwise distances.
\revc{Thus, the algorithm should maintain the following invariant:
\[
    e\in D_t \text{ if and only if } \wit{t}(e)\neq \emptyset. \tag{\dag} \label{invariant}
\]
We show next that the algorithm preserves the invariant when handling each node $t\in V(T)$.
}

\begin{lem}
\label{lem:cqnleaf}
Let $t$ be a leaf node of $T$. Then \revc{Invariant (\ref{invariant})} holds for
\begin{align*}
D_{t}=\{(\alpha_1,\dots,\alpha_k,(d_{i,j})_{1\leq i < j \leq k}) :{}&\alpha_1,\dots,\alpha_k\colon \chi(t)\rightarrow \dom(I), \nonumber \\
&\alpha_1,\dots,\alpha_k\in Q_{t}(I),\nonumber \\  
&  d_{i,j} = \Delta_X(\alpha_i,\alpha_j),1\leq i<j\leq k\}.
\end{align*}
\end{lem}
\begin{proof}
Observe that $\chi(t)=\chi(T_t)$ and $\alpha_1,\dots,\alpha_k$ are the only extensions of $\alpha_1,\dots,\alpha_k$.
Hence, 
\[e=(\alpha_1,\dots,\alpha_k,(d_{i,j})_{1\leq i < j \leq k})\in D_t \iff (\alpha_1,\dots,\alpha_k) \in \wit{t}(e). \qedhere \]
\end{proof}

\begin{lem}
Let $t$ be an introduce node of $T$ which introduces the variable $z$ and let $c$ be its child.
Then, if \revc{Invariant (\ref{invariant})} holds for $D_c$, it also holds for
\begin{align*}
D_{t}=\{(\alpha_1 \cup \beta_1,\dots,\alpha_k \cup \beta_k,(d'_{i,j})_{1\leq i < j \leq k}) :{}&(\alpha_1,\dots,\alpha_k,(d_{i,j})_{1\leq i < j \leq k}) \in D_c, \nonumber \\
&\beta_1,\dots,\beta_k\colon \{z\}\rightarrow \dom(I), \nonumber\\
& \alpha_1\cup \beta_1,\dots, \alpha_k\cup \beta_k \in Q_{t}(I), \nonumber \\
& d'_{i,j} = d_{i,j} +  \Delta_X(\beta_i,\beta_j), 1\leq i < j \leq k \}.
\end{align*}
\end{lem}

\begin{proof}
First notice that $\chi(T_c) \cup \{z\} = \chi(T_t)$ and thus, every literal of $Q_{T_c}$ appears in $Q_{T_t}$.
Furthermore, any variable of $\chi(T_c)$ that appears together with $z$ in a literal has to appear in $\chi(t)$ due to the properties of a tree decomposition.
We can therefore also conclude that a literal appears in $Q_{T_t}$ if and only if it appears in $Q_{T_c}$ or $Q_{t}$.

Now, let the tuple $e=(\alpha_1 \cup \beta_1,\dots,\alpha_k \cup \beta_k,(d'_{i,j})_{1\leq i < j \leq k})$ be in $D_t$ 
and let $e_c=(\alpha_1,\dots,\alpha_k,(d_{i,j})_{1\leq i < j \leq k})$ be a matching tuple in $D_c$.
Thus there is a $(\gamma_1,\dots,\gamma_k)\in \wit{c}(e_c)$.
Importantly, $\gamma_1,\dots,\gamma_k \in I(Q_{T_c}), \alpha_1 \cup \beta_1,\dots, \alpha_k \cup \beta_k \in I(Q_{t})$ and thus, $\gamma_1 \cup \beta_1, \dots, \gamma_k \cup \beta_k \in I(Q_{T_t})$.
Furthermore,  for $1\leq i<j\leq k$, we have:
\begin{align*}
    d'_{i,j}&{} = d_{i,j} + \Delta_X(\beta_i,\beta_j) \\
    &{}= \Delta_X(\gamma_i,\gamma_j) + \Delta_X(\beta_i,\beta_j) \\
    &{} = \Delta_X(\gamma_i \cup \beta_i,\gamma_j \cup \beta_j).
\end{align*}
Thus, $(\gamma_1 \cup \beta_1, \dots, \gamma_k \cup \beta_k) \in \wit{t}(e)$ by definition.

For the reverse direction, let $e=(\alpha_1,\dots,\alpha_k,(d'_{i,j})_{1\leq i < j \leq k})$ be such that there is a $(\gamma_1,\dots,\gamma_k) \in \wit{t}(e)$.
Thus, we can immediately conclude that $\gamma_1|_{\chi(t)},\dots,\gamma_k|_{\chi(t)}\in I(Q_{t})$ while $\gamma_1|_{\chi(T_c)},\dots,\gamma_k|_{\chi(T_c)}\in I(Q_{T_c})$.
Furthermore,  for $1\leq i<j\leq k$, we have:
\begin{align*}
    d'_{i,j} - \Delta_X(\gamma_i|_{\{z\}},\gamma_j|_{\{z\}}) {}&= \Delta_X(\gamma_i,\gamma_j) - \Delta_X(\gamma_i|_{\{z\}},\gamma_j|_{\{z\}}) \\
    &{} = \Delta_X(\gamma_i|_{\chi(T_t)},\gamma_j|_{\chi(T_t)}).
\end{align*}
Thus, 
\[e_c=(\gamma_1|_{\chi(c)} ,\dots,\gamma_k|_{\chi(c)},(d'_{i,j}-\Delta_X(\gamma_i|_{\{z\}},\gamma_j|_{\{z\}}))_{1\leq i < j \leq k})\in D_{c}.\] 
as $(\gamma_1|_{\chi(T_c)},\dots,\gamma_k|_{\chi(T_c)})\in \wit{c}(e_c)$.
Defining $\beta_1=\gamma_1|_{\{z\}},\dots, \beta_k=\gamma_k|_{\{z\}}$ then ensures that 
$(\gamma_1|_{\chi(T_c)} \cup \beta_1,\dots,\gamma_k|_{\chi(T_c)} \cup \beta_k,(d'_{i,j})_{1\leq i < j \leq k}) = (\alpha_1,\dots,\alpha_k,(d'_{i,j})_{1\leq i < j \leq k})$ is in $D_t$.
\end{proof}

\begin{lem}
Let $t$ be a forget node of $T$ and $c$ its child. 
Then, if \revc{Invariant (\ref{invariant})} holds for $D_c$, it also holds for
\begin{align*}
    D_{t}=\{(\alpha_1|_{\chi(t)},\dots,\alpha_k|_{\chi(t)},(d_{i,j})_{1\leq i < j \leq k}) :{}&(\alpha_1,\dots,\alpha_k,(d_{i,j})_{1\leq i < j \leq k})\in D_c\}.
\end{align*}
\end{lem}

\begin{proof}
Let $z$ be the forgotten variable.
The claim follows from the fact that $Q_{T_t}=Q_{T_c}$ and $\chi(t)\subseteq\chi(c)$, and thus, 
\[\wit{t}(\alpha'_1,\dots,\alpha'_k,(d_{i,j})_{1\leq i < j \leq k})=\bigcup_{\substack{\beta_i\colon \{z\}\rightarrow \dom(I)\\ l=1,\dots,k}} \wit{c}(\alpha'_1\cup \beta_1,\dots,\alpha'_k\cup \beta_k,(d_{i,j})_{1\leq i < j \leq k}). \qedhere\]
\end{proof}

\begin{lem}
\label{lem:cqnjoin}
Let $t$ be a join node of $T$ with children $c_1$ and $c_2$.
Then, if \revc{Invariant (\ref{invariant})} holds for $D_{c_1}$ and $D_{c_2}$, it also holds for
\begin{align*}
    D_{t}=\{(\alpha_1,\dots,\alpha_k,(d_{i,j})_{1\leq i < j \leq k}) :{}&(\alpha_1,\dots,\alpha_k,(d'_{i,j})_{1\leq i < j \leq k}) \in D_{c_1}, \nonumber \\
    &(\alpha_1,\dots,\alpha_k,(d''_{i,j})_{1\leq i < j \leq k}) \in D_{c_2}, \nonumber \\
    & d_{i,j} = d'_{i,j} + d''_{i,j} -  \Delta_X(\alpha_i,\alpha_j), 1\leq i < j \leq k \}.
\end{align*}
\end{lem}

\begin{proof}
First notice that $\chi(T_{c_1})\cup \chi(T_{c_2})=\chi(T_t)$ and thus, literals that appear in $Q_{T_{c_1}}$ or $Q_{T_{c_2}}$ also appear in $Q_{T_t}$.
Moreover, if two variables appear in the same literal in $Q_{T_t}$ but they no longer jointly 
occur in $\chi(t)$, then these variables have to appear together in either $T_{c_1}$ or $T_{c_2}$.
We can, therefore, observe that a literal appears in $Q_{T_t}$ if and only if it appears in 
$Q_{T_{c_1}}$ or $Q_{T_{c_2}}$.

We start with some 
\[ 
    e_1=(\alpha_1,\dots,\alpha_k,(d'_{i,j})_{1\leq i < j \leq k}) \in D_{c_1} \text{ and } 
    e_2=(\alpha_1,\dots,\alpha_k,(d''_{i,j})_{1\leq i < j \leq k}) \in D_{c_2}. \]
Now let $(\gamma'_1,\dots,\gamma'_k)\in \wit{c_1}(e_1)$ and $(\gamma''_1,\dots,\gamma''_k)\in \wit{c_2}(e_2)$ witness this, respectively.
By the above observation, $\gamma'_1\cup \gamma''_1, \dots, \gamma'_k\cup \gamma''_k  \in Q_{T_t}(I)$ and, for $1\leq i < j \leq k$, we have: 
\begin{align*}
    \Delta_X(\gamma'_i\cup \gamma''_i,\gamma'_j\cup \gamma''_j) &{}= \Delta_X(\gamma'_i,\gamma'_j) + \Delta_X(\gamma''_i, \gamma''_j) - \Delta_X(\gamma'_i\cap \gamma''_i,\gamma'_j\cap \gamma''_j) \\
    &{}= d'_{i,j} + d''_{i,j} - \Delta_X(\alpha_i, \alpha_j).
\end{align*}
Hence, $e=(\alpha_1,\dots,\alpha_k,(d'_{i,j} + d''_{i,j} - \Delta_X(\alpha_i,\alpha_j))_{1\leq i < j \leq k})\in D_{t}$ is justified as $(\gamma'_1\cup \gamma''_k,\dots,\gamma'_1\cup \gamma''_k)\in \wit{t}(e)$.

Conversely, let $e=(\alpha_1,\dots,\alpha_k,(d_{i,j})_{1\leq i < j \leq k})$ be a tuple such that there is a $(\gamma_1,\dots,\gamma_k)\in \wit{t}(e)$.
Thus, we can immediately conclude that the restrictions $\gamma_1|_{\chi(T_{c_1})},\dots,\gamma_k|_{\chi(T_{c_1})}$ are in~$I(Q_{T_{c_1}})$ while the restrictions $\gamma_1|_{\chi(T_{c_2})},\dots, \gamma_k|_{\chi(T_{c_2})}$ are in $I(Q_{T_{c_2}})$.
This implies that
\begin{align*}
(\gamma_1|_{\chi(t)},\dots,\gamma_k|_{\chi(t)}&,(\Delta_X(\gamma_i|_{\chi(T_{c_1})},\gamma_k|_{\chi(T_{c_1})}))_{1\leq i < j \leq k})\in D_{c_1}, \\
(\gamma_1|_{\chi(t)},\dots,\gamma_k|_{\chi(t)}&,(\Delta_X(\gamma_i|_{\chi(T_{c_2})},\gamma_k|_{\chi(T_{c_2})}))_{1\leq i < j \leq k}) \in D_{c_2}.
\end{align*}
Lastly, we can compute for $1\leq i < j \leq k$:
\begin{align*}
    d_{i,j}&{}= \Delta_X(\gamma_i,\gamma_j) \\
    &{}= \Delta_X(\gamma_{i}|_{\chi(T_{c_1})},\gamma_j|_{\chi(T_{c_1})}) + \Delta_X(\gamma_{i}|_{\chi(T_{c_2})},\gamma_j|_{\chi(T_{c_2})}) - \Delta_X(\gamma_{i}|_{\chi(t)},\gamma_j|_{\chi(t)}),
\end{align*}
implying that $e\in D_t$.
\end{proof}
Importantly, Lemmas \ref{lem:cqnleaf} through \ref{lem:cqnjoin} ensure that after the bottom-up traversal of 
\autoref{algo:two}, \revc{Invariant (\ref{invariant})} is satisfied for $D_r$.
We now show that the algorithm correctly determines  from $D_r$
if there is a diversity set of size $k$ which has diversity 
exceeding $d$.

\begin{lem}
\label{lem:cqn-finalization}
If \revc{Invariant (\ref{invariant})} holds for $D_r$ then there exist solutions $\{\gamma_1,\dots,\gamma_k\}\subseteq Q(I)$ with $\delta(\gamma_1,\dots,\gamma_k)\geq d$ if and only if there is a tuple $(\alpha_1,\dots,\alpha_k,(d_{i,j})_{1\leq i < j \leq k}) \in D_r$ such that $d_{i,j}> 0, 1\leq i < j \leq k$ and  $f((d_{i,j})_{1\leq i < j \leq k})\geq d$
\end{lem}
\begin{proof}
First observe that $Q=\exists Y Q_{T_r}$ and thus, $Q(I)= \{\gamma|_X : \gamma \in Q(T_r)\}$.
Furthermore, for $\gamma,\gamma'\in Q(T_r)$ we have $\Delta(\gamma|_X,\gamma|_X)=\Delta_X(\gamma,\gamma')$.

Now assume $k$ solutions $\{\gamma_1|_X,\dots,\gamma_k|_X\}\subseteq Q(I)$ with $\delta(\gamma_1|_X,\dots,\gamma_k|_X)\geq d$ to exist.
Consider the tuple $e=(\gamma_1|_{\chi(r)},\dots,\gamma_k|_{\chi(r)},(\Delta_X(\gamma_i,\gamma_j))_{1\leq i < j \leq k}).$
By definition, we have $(\gamma_1,\dots,\gamma_k)\in\wit{t}(e)$ and hence, $e\in D_r$.
Furthermore, $\Delta_X(\gamma_i,\gamma_j) = \Delta(\gamma_i|_X,\gamma_j|_X) > 0$ for $1\leq i < j \leq k$ and 
\[f((\Delta_X(\gamma_i,\gamma_j))_{1\leq i < j \leq k}) = f((\Delta(\gamma_i|_X,\gamma_j|_X))_{1\leq i < j \leq k}) = \delta(\gamma_1|_X,\dots,\gamma_k|_X)\geq d.\]

For the reverse direction assume such a tuple $e=(\alpha_1,\dots,\alpha_k,(d_{i,j})_{1\leq i < j \leq k}) \in D_r$ with $d_{i,j}> 0,1\leq i < j \leq k,$ and  $f((d_{i,j})_{1\leq i < j \leq k})\geq d$ to exist.
Thus, there also exists $(\gamma_1,\dots,\gamma_k)\in \wit{r}(e)$ and we claim that $\{\gamma_1|_X,\dots,\gamma_k|_X\}$ is a diversity set as required.
First observe that $\gamma_i|_X$ is different to $\gamma_j|_X$ for $1\leq i < j \leq k$ as $\Delta(\gamma_i|_X,\gamma_j|_X)= \Delta_X(\gamma_i,\gamma_j) = d_{i,j} > 0$.
Secondly, $\delta(\gamma_1|_X,\dots,\gamma_k|_X) = f((\Delta(\gamma_i|_X,\gamma_j|_X))_{1\leq i < j \leq k}) = f((\Delta_X(\gamma_i,\gamma_j))_{1\leq i < j \leq k}).$
\end{proof}

We now show that the bottom-up step is also possible in the required time bound.

\begin{lem}
\label{lem:cqn-runtime}
Let $t$ be an arbitrary node in $T$, $\omega$ the width of the tree decomposition.
Then, $D_t$ can be computed in time $\O(\dom(I)^{2\cdot k\cdot (\omega + 1)}\cdot (|X|+1)^{k(k-1)} \cdot(|Q| + k^2\cdot \omega))$ given the sets of its children.
\end{lem}
\begin{proof}
This running time is achieved by a naive implementation.
For a leaf node, we simply have to iterate through the $|\dom(I)|^{|\chi(t)|\cdot k}$ options for $\alpha_1,\dots,\alpha_k\colon\chi(t)\rightarrow \dom(I)$, check that all assignments are answers to $Q_{t}$ (each relevant table of $I$ is at most of size $\dom(I)^{\omega + 1}$), and compute the Hamming distances.

For an introduce node, we iterate through all elements of $D_c$ and all $|\dom(I)|^k$ possibilities for $\beta_1,\dots,\beta_2\colon \{z\}\rightarrow \dom(I)$, again check whether all assignments are answers to $Q_{t}$, and update the distances.
Note that the size of the set $D_c$ is at most $|\dom(I)|^{k\cdot (\omega+1)}\cdot (|X|+1)^{\frac{k(k-1)}{2}}$ by definition since $|X|$ is an upper bound on the pairwise Hamming distance.

For a forget node, we simply need to perform a projection.
Lastly, for a join node, we iterate through $D_{c_1}\times D_{c_2}$, check whether the assignments $\alpha_1,\dots,\alpha_k$ match, and then update the distances.

All of the cases can therefore clearly be handled in the given time bound.
\end{proof}

\noindent
With this, we can now show Theorem \ref{theorem:cqnXP}.

\begin{proof}[Proof of \autoref{theorem:cqnXP}]
To prove the result, we only have to ensure that finding a suitable tree decomposition and applying the algorithm 
is possible in the required time bound as the correctness of this procedure follows directly from 
Lemmas \ref{lem:cqnleaf} to \ref{lem:cqn-finalization}.

Due to
\cite{DBLP:journals/siamcomp/Bodlaender96} and \cite{DBLP:books/sp/Kloks94} computing a width optimal nice tree decomposition is possible in linear time.
Furthermore, the number of nodes in this tree decomposition is linear in $|Q|$ and, thus, performing the bottom-up traversal of the algorithm is possible in the required time bound due to Lemma \ref{lem:cqn-runtime}.
Lastly, we also have to look at the running time of the final step of the algorithm.
There, we possibly have to evaluate $f$ for each tuple in $D_r$.
But, since $|D_r|\leq \dom(I)^{k(\omega + 1)}\cdot (|X|+1)^\frac{k(k-1)}{2}$ and $f$ is computable in polynomial time (in $k$ and $|X|$), also this step is possible in the required time bound.
\end{proof}

We conclude this section by again stressing the analogy with \autoref{algo:one} for ACQs: 
First, we have omitted from our description of \autoref{algo:two} how to compute a concrete witnessing 
diversity set in the case of a yes-answer.
This can be done exactly as in \autoref{algo:one} by maintaining the same kind of provenance information. 
And second, it is possible to speed up the present algorithm by applying the same kind of considerations as in 
Section~\ref{sec:speedups}. 
It is thus possible to reduce the query complexity to $\fpt$ for the diversity measure $\deltasum$ and even further to $\ptime$ if we allow duplicates in the diversity set.

\section{Conclusion and Future Work}
\label{sec:conclusion}

In this work, we have had a fresh look at the \Diverse problem of query answering. 
For CQs and extensions thereof, we have proved a collection of complexity results, both for the parameterized and the non-parameterized case. 
To get a chance of reaching tractability or at least fixed-parameter tractability (when considering the size $k$ of the diversity set as the parameter), we have restricted ourselves to acyclic CQs and CQs with negation of 
bounded treewidth, respectively. 
For the chosen settings, our complexity results are fairly complete. 
The most obvious gaps left for future work are concerned with the query complexity of ACQs and CQs with negation of bounded treewidth. 
For the parameterized case, we have $\xp$-membership but no fixed-parameter intractability result in the form of $\wone$-hardness. 
And for the non-parameterized case, it is open if the problems are also $\np$-hard as we have shown for the data complexity. 

It should be noted that the restriction to acyclic CQs is less restrictive than it may seem at first glance.
\revc{As was mentioned in Section \ref{sec:preliminaries}, 
query evaluation of CQs with 
bounded hypertree width (likewise, CQs with bounded generalized or fractional 
hypertree width) can be efficiently reduced to query evaluation of acyclic CQs.
Hence}, 
our upper bounds (in particular, the $\xp$- and $\fpt$-membership results in Section~\ref{sec:CQs})
are easily generalized to CQs of bounded hypertree-width~\cite{DBLP:journals/jcss/GottlobLS02}. Moreover, 
recent  empirical studies of millions of queries from query logs 
\cite{DBLP:journals/vldb/BonifatiMT20}
and thousands of queries from benchmarks
\cite{DBLP:journals/jea/FischlGLP21}
have shown that CQs typically have hypertree-width at most 3.

\revc{
A yet more powerful
width measure than $\hw$, $\ghw$, and $\fhw$
is submodular-width ($\smw$) introduced in \cite{DBLP:journals/jacm/Marx13}. In fact, Marx showed that bounded $\smw$, in a sense, 
exactly characterizes the class of CQs for which query evaluation is 
fixed-parameter tractable, parameterized by the size of the query. 
In contrast
to the other width measures, there is no obvious extension of our results to CQs of bounded $\smw$. Indeed,
query evaluation based on bounded $\smw$ works by partitioning 
the database via so-called ``heavy-light splitting''  
(i.e., treating attribute values with high vs. low frequency differently). 
Clearly, this reduces the problem of CQ-evaluation to a 
problem of UCQ-evaluation and we have seen in 
Section \ref{sec:UCQs} that the diversity problem for UACQs immediately leads 
to intractability. 
Hence, a completely different approach would be needed to extend our results to 
CQs with bounded $\smw$.
}

\revc{
For CQs with negation, a stronger restriction than acyclicity (or bounded hw, ghw, fhw) is needed to achieve
tractability of query evaluation \cite{DBLP:journals/tcs/Lanzinger23}. 
Consequently, we have studied CQs with negation of bounded treewidth. Another
interesting restriction on the structure of CQs with negation to achieve tractable query evaluation is 
$\beta$-acyclicity \cite{DBLP:conf/csl/Brault-Baron12,DBLP:conf/pods/NgoNRR14,%
DBLP:conf/icdt/CapelliI24} or, more generally, bounded nest-set width 
\cite{DBLP:journals/tcs/Lanzinger23,DBLP:conf/icdt/CapelliI24}. 
Both these properties are defined 
via a form of variable elimination. In contrast to acyclic queries or queries of bounded tw (and, likewise, 
bounded hw, ghw, fhw), there is no form of {\em decomposition} 
to characterize $\beta$-acyclicity or bounded 
nest-set width. Hence, there is no obvious way how to extend our results on CQs with negation to
$\beta$-acyclic queries or queries with bounded nest-set width. 
We leave both, the diversity study of 
CQs with bounded $\smw$ and 
of CQs with negation of bounded nest-set width as interesting questions for future work.  
}

Another direction for future work is motivated by 
a closer look at our $\fpt$- and $\xp$-membership results: even though such parameterized complexity results are generally considered as favorable (in particular, $\fpt$), the running times are exponential in the parameter~$k$. 
As we allow larger values of $k$, these running times may not be acceptable anymore. It would therefore be interesting
to study the diversity problem also from an approximation point of view -- in particular, 
contenting oneself with an approximation of the desired diversity. 

A further modification of our settings is related to 
the choice of a different
distance measure between two answer tuples and different aggregators. As far as the distance measure is concerned, 
we have so far considered data values as untyped and have therefore studied only the Hamming distance between tuples. 
For numerical values, one might of course take the difference between values into account. More generally, 
one could consider a metric on the domain, which then induces a metric on tuples that can be used as a distance measure. As far as the aggregator is concerned, we note that most of our upper bounds apply to arbitrary 
(polynomial-time computable) aggregate functions. 
On the other hand, our lower bounds hold for any (polynomial-time computable) ws-monotone aggregate functions. 
This seems quite a natural choice as almost all natural aggregators in this setting -- including sum and min
-- are ws-monotone. 
A problem 
strongly related to \Diverse is \Similar \cite{DBLP:journals/tplp/EiterEEF13}, where one is interested in finding solutions close to each other. 
Parts of our approach can naturally be adapted to this setting but we leave the in-depth study of  
\Similar for future work.

\bibliographystyle{alphaurl}

\bibliography{lmcs}

\end{document}

%% file: myMacros.tex
\renewcommand{\phi}{\varphi}
\renewcommand{\O}{\mathcal{O}}
\newcommand{\var}{\mathit{var}}

\newcommand{\dom}{\mathit{dom}}

\newcommand{\ext}[1]{\mathit{ext}_{#1}\xspace}
\newcommand{\wit}[1]{\mathit{wit}_{#1}\xspace}

\newcommand{\hw}{\mathit{hw}}
\newcommand{\ghw}{\mathit{ghw}}
\newcommand{\fhw}{\mathit{fhw}}
\newcommand{\tw}{\mathit{tw}}
\newcommand{\smw}{\mathit{smw}}

\newcommand{\cq}{CQ\xspace}
\newcommand{\cqn}{CQ$^\lnot$\xspace}
\newcommand{\cqsn}{CQs$^\lnot$\xspace}
\newcommand{\acq}{ACQ\xspace}
\newcommand{\cqs}{CQs\xspace}

\newcommand{\setdefsep}{:}

\newcommand{\Iold}{I_{\mathit{old}}}
\newcommand{\Inew}{I_{\mathit{new}}}
\newcommand{\Istar}{I^*}

\newcommand{\IndependentSet}{\textsc{Independent Set}\xspace}

\newcommand{\Similar}{\textsf{Similarity}\xspace}
\newcommand{\Diverse}{\textsf{Diversity}\xspace}
\newcommand{\DiverseQ}{\textsf{Diverse-}$\mathcal{Q}$\xspace}
\newcommand{\DiverseQsum}{\textsf{Diverse}$_{\mathsf{sum}}$-$\mathcal{Q}$\xspace}

\newcommand{\DiverseQmon}{\textsf{Diverse}$_{\mathsf{mon}}$-$\mathcal{Q}$\xspace}
\newcommand{\DiverseQstrmon}{\textsf{Diverse}$_{\mathsf{wsm}}$-$\mathcal{Q}$\xspace}

\newcommand{\DiverseFOmon}{\textsf{Diverse}$_{\mathsf{mon}}$-\textsf{FO}\xspace}

\newcommand{\DiverseCQn}{\textsf{Diverse-CQ}$^\lnot$\xspace}

\newcommand{\DiverseCQnstrmon}{\textsf{Diverse}$_{\mathsf{wsm}}$-\textsf{CQ}$^\lnot$\xspace}

\newcommand{\DiverseACQ}{\textsf{Diverse-ACQ}\xspace}
\newcommand{\DiverseACQsum}{\textsf{Diverse}$_{\mathsf{sum}}$-\textsf{ACQ}\xspace}

\newcommand{\DiverseACQmon}{\textsf{Diverse}$_{\mathsf{mon}}$-\textsf{ACQ}\xspace}
\newcommand{\DiverseACQstrmon}{\textsf{Diverse}$_{\mathsf{wsm}}$-\textsf{ACQ}\xspace}

\newcommand{\DiverseUACQ}{\textsf{Diverse-UACQ}\xspace}

\newcommand{\DiverseUACQstrmon}{\textsf{Diverse}$_{\mathsf{wsm}}$-\textsf{UACQ}\xspace}

\newcommand{\deltasum}{\delta_{\mathsf{sum}}}

\newcommand{\deltamon}{\delta_{\mathsf{mon}}}
\newcommand{\deltastrmon}{\delta_{\mathsf{wsm}}}

\newcommand{\CQ}{\mathsf{CQ}}
\newcommand{\UCQ}{\mathsf{UCQ}}
\newcommand{\CQneg}{\mathsf{CQ}^\neg}
\newcommand{\FO}{\mathsf{FO}}
\newcommand{\ACQ}{\mathsf{ACQ}}
\newcommand{\UACQ}{\mathsf{UACQ}}

\newcommand{\sols}[2]{\ensuremath{#1(#2)}} % #1: Query/Atom/... ; #2: Instance

\newcommand{\ptime}{\mathsf{P}}
\newcommand{\np}{\mathsf{NP}}
\newcommand{\fpt}{\mathsf{FPT}}
\newcommand{\xp}{\mathsf{XP}}

\newcommand{\wone}{\mathsf{W}[1]}
\newcommand{\nop}[1]{}

\newcommand{\pol}{\ensuremath{\mathit{poly}}}

\newcommand{\calQ}{\mathcal{Q}\xspace}

\newcommand{\Dnew}{D^{\textit{new}}}